\documentclass[a4paper,fleqn,usenatbib]{mnras}

\usepackage[T1]{fontenc}
\usepackage{ae,aecompl}

\usepackage{graphicx}	
\usepackage{amsmath}	
\usepackage{amssymb}	
\title[Synchronized Globular Cluster Formation]{On Synchronized Globular Cluster Formation over Supra-galactic Scales: A Virgo-Centaurus Group Connection}
\author[J.C. Forte]{
Juan C. Forte,$^{1,2}$\thanks{E-mail: planeta.jcf@gmail.com}
\\
$^{1}$Academia Nacional de Ciencias de Buenos Aires, Av. Alvear 1711, 1014, Buenos Aires, Argentina\\
$^{2}$Academia Nacional de Ciencias, Av. Velez Sarsfield 249, X5000JJC, Córdoba, Argentina\\
}
\date{Accepted XXX. Received YYY; in original form ZZZ}
\pubyear{2022}
\begin{document}
\label{firstpage}
\pagerange{\pageref{firstpage}--\pageref{lastpage}}
\maketitle
\begin{abstract}
This work reports the detection of a multi-peaked colour pattern in the integrated colours distribution of globular clusters associated to the giant elliptical galaxy $NGC~4486$, using $Next~Generation~Virgo~Survey$ data. This feature is imprinted on the well known $bimodal$ colour distribution of these clusters. Remarkably, the pattern is similar to that found in previous works, based on photometry from the $HST-Advanced~Camera~Virgo~Survey$, in less massive Virgo galaxies. This characteristic can be traced up to to 45\arcmin~($\approx$217 $Kpc$) in galactocentric radius. This suggests that globular cluster formation in Virgo has been regulated, at least partially, by a collective process composed by several discrete events, working on spatial scales comparable to the size of the galaxy cluster.
 Furthermore, the presence of a similar colour pattern in $NGC~5128$, at the outskirsts of the $Virgo~Super-cluster$, poses an intriguing question about the spatial scale of the phenomenon. The nature of the process that leads to the colour pattern is unknown. However, energetic events connected with galaxy or sub-galaxy cluster mergers and $SMBH$ activity, in the early Universe, appear as possible candidates to explain an eventual enhancement/quenching of the globular clusters formation, reflected in  the modulation of their integrated colours. Such events, presumably, may also have had an impact on the whole star formation history in Virgo galaxies.
\end{abstract}
\begin{keywords}
galaxies: star clusters: general

\end{keywords}
\section{Introduction}
\label{Intro}
 An increasing volume of observations in the high redshift domain \citep[see, for example,][]{Izumi2021}, as well as inferences from cosmological models, converge to show that the early Universe was dominated by complex and energetic interactions that had an impact on the post temporal evolution of galaxies. 
 
 In turn, and for a long time, globular clusters ($GCs$) have been known as old systems and recent estimates place the origins of the oldest ones within a couple of hundred $Myr$ after the Big Bang \citep[e.g.][]{Valcin2020}. In this context, some $GCs$ that are observable in the the local Universe, appear as potential candidates to search for eventual information about those early events. 
 
 For the last twenty five years, a dominant paradigm in extra galactic $GCs$ research, has been the so called ''bimodality''. This term indicates that $\bf{most}$~$GCs$ belong to one of two families. On one side, $blue$ clusters with low metallicity and shallow spatial distributions and, on the other, $red$ clusters with higher metallicities and more concentrated towards the galaxy centres.
  
 Written several years ago, the dawn of this concept has been thoroughly described in \citet{Brodie2006}. A connected, and even earlier precedent, can be found in \citet{Forte1982} who explained previous results by \citet{Strom1981} in $NGC ~4486$ on the basis of  ''original'' (associated to the galaxy halo) and ''external'' (accreted from Virgo dwarf galaxies) $GCs$ . 
 
 Integrated colours of old stellar systems are a proxy to metallicity and the $GCs$ colour distributions ($GCCDs$) give an idea about the chemical content of each of the cluster sub-populations. Bimodality seems to have a direct connection with the halo-bulge structure observed in the $MW$ and in other galaxies. A key contribution to this subject, based on $HST-ACSVS$ observations of $GCs$ in the  $Virgo~ cluster$, has been presented by \citet{Peng2006}.
 
 The interest in the study of $GCCDs$ in different galaxies had a strong motivation in the frame of the \citet{Ashman1992} scenario. These authors suggested that merger events may have triggered the formation of massive stellar clusters  (eventually, $GC~progenitors$) then leaving distinct imprints on the $GCCDs$. However, the attention of following papers has been more concentrated on the bimodality issue than on the eventual presence of subtle features. Deviations from bimodality are usually explained as the result of the sampling noise associated with stochastic processes.
 
 A summary about the status of bimodality in the context of extragalactic $GCs$ research was presented some years ago in \citet{Harris2017}, who also emphasized the need to advance beyond that concept. 
 
 An exploratory attempt in that direction was introduced in \citet{Forte2017} ($F2017$ in what follows). The idea behind that work is that, even though the $GCCDs$ in different galaxies should reflect their individual formation process, the presence of common features might be detectable on large composite samples of $GCs$. This approach was tested on the $HST-ACSVS$ photometric data published by \citet{Jordan2009} and \citet{Jordan2015} for $GC$ systems in the $Virgo$ and $Fornax$ clusters, respectively (see Section 2).
  
 The presence of multi-peaked colour patterns was detected on composite $GC$ samples associated with non-giant, moderately bright galaxies (i.e., $M_{g}$ from -20.2 to -19.2), in both galaxy clusters. These features were dubbed as the $template~Virgo$ and $template~Fornax~patterns$~($TVP$ and $TFP$, respectively). 
 
 A comparison of the $GC$ colour patterns in these galaxy clusters shows some similarities and differences. For example, the Virgo $GCs$ exhibit two blue components, with peaks at $(g-z)=$ 0.85 and 0.95, while the Fornax pattern has a single broad blue peak at $(g-z)$=0.94. Other five, redder colour peaks in Fornax, are some 0.03 mag bluer than their Virgo counterparts.

 A subsequent paper, \citet{Forte2019}, extended the analysis to the central regions of giant galaxies in Virgo and confirmed the presence of $TVP$-like colour structures in the inner regions of these systems.
 
 Photometric errors, field contamination, and statistical noise, were analysed and rejected as possible spurious origins of these patterns. 
  
 The physical origin of the template patterns is not known and further efforts in clarifying their nature might not be justified without a tentative scenario. A (cautious) working hypothesis, presented in $F2017$, assumes that colour peaks in the $GCCDs$ are set during  periods of star-burst activity, including an enhancement of $GC$ formation, within a total temporal span of $\approx$1.5~$Gyr$~in the time scale defined by the $SSP$ models presented by \citet{Bressan2012}. 
 
 If the colour patterns arise as the result of different external events, a chronological scale could be defined since chemical abundance $[Z/H]$ correlates with time. In fact the $GCCD$ may reflect either a temporal sequence (where $[Z/H]$ is a $clock$), a positional sampling (halo; bulge) , or eventually, a combination of  both. This seems to be the case of $GCs$ in the MW \citep{Leaman2013, Massari2019}, that exhibits a bifurcated age-metallicity relation (i.e. halo/blue or bulge/red $GCs$). 
 
 If comparable relations hold in other galaxies, different events may have left simultaneous imprints (colour peaks) on both the blue and red $GC$ populations, as their respective metallicities increased simultaneously with time (see Fig. 30 in $F2017$). The hypothesis requires external stimuli, working on large spatial scales, and able to unleash simultaneous $GC$ formation on all cluster galaxies. It is worth mentioning that, evidence in favor of $GC$ formation synchronicity, in the case of the $MW$ and the $LMC$, has been presented by \citet{Wagner-Kaiser2017}.
 
 Some results that deserve further attention in connection with the tentative landscape described above are:
\begin{enumerate}
\item The existence of intense star-formation widespread on a spatial scale of several $Mpc$ can be seen in the low redshift ($z=$0.19) ''Sausage cluster'' ($CIZAJ22428+5301$). In this merger of galaxy clusters \citet{Stroe2015}, and \citet{Sobral2015}, find a gravitational shock wave, fuelled by the merger, that has triggered a viral star forming process within a short time interval.
\item \citet{Forrest2020} analyse the characteristics of galaxies with a range of redshift $z$ from 3 to 6 and conclude that their star formation histories are compatible with the presence of several, intense and short lived, star formation events, followed by quenching periods. This is consistent, for example, with a previous analysis by \citet{Schulz2015} in their study of the stellar populations in $NGC~1399$, one of the central galaxies in the $Fornax~cluster$.
\item \citet{Navarro2021} find evidence of the impact of super massive black holes $(SMBHs)$ on modulating the star formation rates in satellite galaxies. The main argument behind their results is that quiescent satellite galaxies are less frequent in the direction of the minor axis of their central galaxies. This is interpreted as a consequence of the effect of the $SMBHs$ activity on the circumgalactic medium and the star formation rates. 
 \end{enumerate}
 All these results may provide a starting frame to clarify which physical mechanism, or eventually a combination of them, could be responsible for the observed template patterns. For example, merger induced star bursts, followed by a period of quenching originated by $SMBH$ feedback, looks as a possible scenario for further exploration.
 
 This work is focused on the abundant $GC$ population of the giant elliptical $NGC~4486$, one of the central galaxies in the $Virgo~ cluster$,  in an effort to detect the eventual presence of a colour pattern comparable to that seen in the non-giant galaxies in the same cluster. The detection of a $TVP-like$ structure in the inner region of $NGC~4486$ ($R_{gal}=$0.6\arcmin~to 1.5\arcmin) has been already noted in \citet{Forte2019} (see their Fig. 7) using $ACSVS$ data \citep{Jordan2009}. The eventual detection of such a pattern on a large angular field, and on a different data set, would provide a strong support in favor of a physical nature of the observed $GCs$ colour modulation.
 
 For the following discussion we adopted the photometric catalogue presented by \citet{Oldham2016} ($OA2016$ in what follows), based on $ugriz$~$NGVS$ data \citep{Ferrarese2012}. Those authors performed a census of the $NGC~4486$~$GC$ populations, based on a Bayesian approach. Their model, aimed at describing the overall properties of the $GC$ system, includes twenty three parameters, and assigns to each object a given probability of being a $GC$ or a field object, based on its image shape, brightness, colour and spatial position.
 
 As an alternative, and complementary procedure, the present work separates $GCs$ and field contaminants through fits of $ugriz$ pseudo-continuums to the photometric data of each object \citep{Forte2013}, as explained in the Appendix.
 
 This work also introduces a random sub-sampling ($RSS$) approach, similar to that used in $Bootstrapping$ analysis, aiming at identifying colour structures in the $GCCDs$, that can be compared with those found in $F2017$. 
 
  In this approach, sub-samples of a larger $GC$ population, are randomly and repetitively chosen. Colour peaks on each sub-sample are identified on their smoothed $GCCDs$ (at the colours with null derivatives) and catalogued. 
 
 We note that the $RSS$ analysis allows for the detection of colour peaks possibly hidden in colour intervals where the $GCCDs$ exhibit high slopes (e.g., the blue and red wings of the dominant blue $GCs$ sub-population).
 
 The structure of the paper is as follows: Section 2 presents a review of the Virgo pattern, introducing a $RSS$ analysis of composite $GC$ samples. Section 3 deals with the homogenization of the colour-colour relations relevant to this work (as detailed in the Appendix). The results from the analysis of the $ugriz$ photometric data are given in Section 4. The characteristics of the colour pattern found in the field of $NGC~4486$ are discussed in Section 5. The case of $GCs$ associated to $NGC~5128$, at the edge of the $Virgo Super-cluster$, is presented in Section 6. Finally, a summary of the results and the conclusions are presented in Sections 7 and 8, respectively.
\section{Reviewing the Template Virgo Pattern}
\label{sec2}
\begin{figure}
\vspace{-0.5cm}
	\includegraphics[width=\columnwidth]{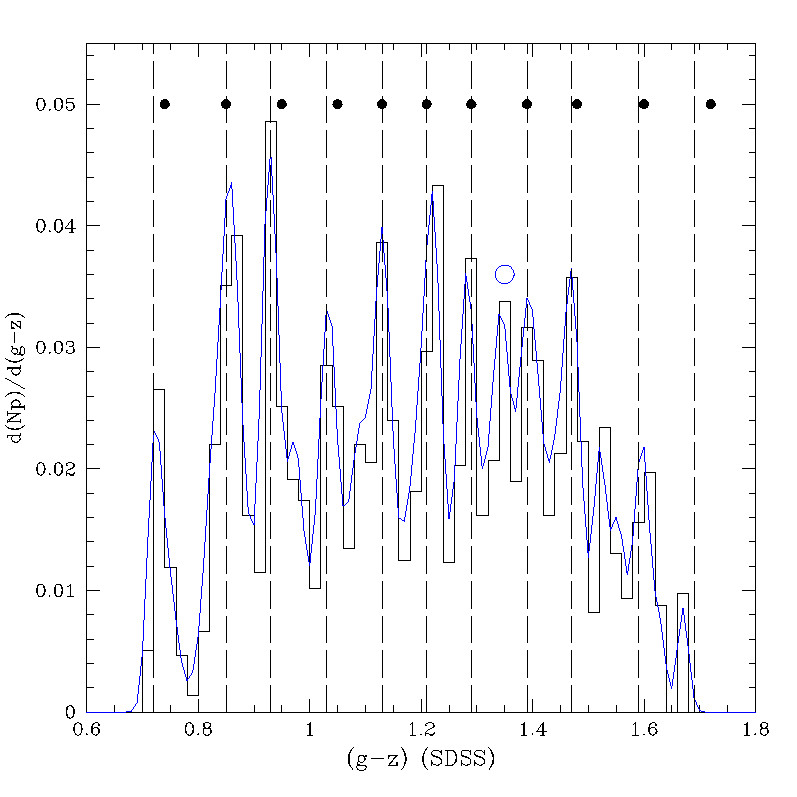}
    \caption{$(g-z)$ colour peak frequencies derived from a $RSS$ routine for 2235 $GCs$ colours in seventy one Virgo non-giant galaxies.
    The histogram has 0.02 mag bins, while the blue line corresponds to a Gaussian smoothing kernel with 0.023 mag $FWHM$. The open dot at $(g-z)$=1.35 indicates a peak not detected in a previous analysis (see text). Dashed lines correspond to the revised $TVP$ colours.
}
    \label{fig:fig1}
\end{figure}
\begin{figure}
\vspace{-0.5cm}
	\includegraphics[width=\columnwidth]{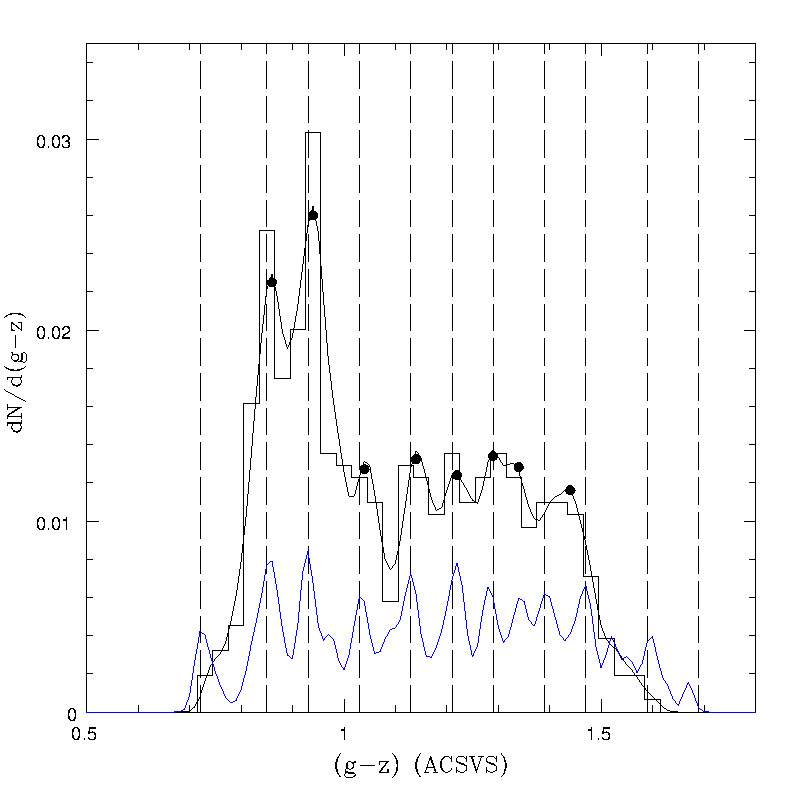}
    \caption{$(g-z)$ colour distribution corresponding to a composite sample of 511 clusters in six galaxies with $M_{g}$ from -20.2 to -19.7. The blue line is the (arbitrarily) scaled down frequency curve shown in Fig. 1. The peak at $(g-z)$$\approx$ 1.35 only appears for clusters in $NGC~4442$. Dashed lines correspond to the $TVP$ colours.
} 
    \label{fig:fig2}
\end{figure}
\begin{figure}
\vspace{-0.5cm}
	\includegraphics[width=\columnwidth]{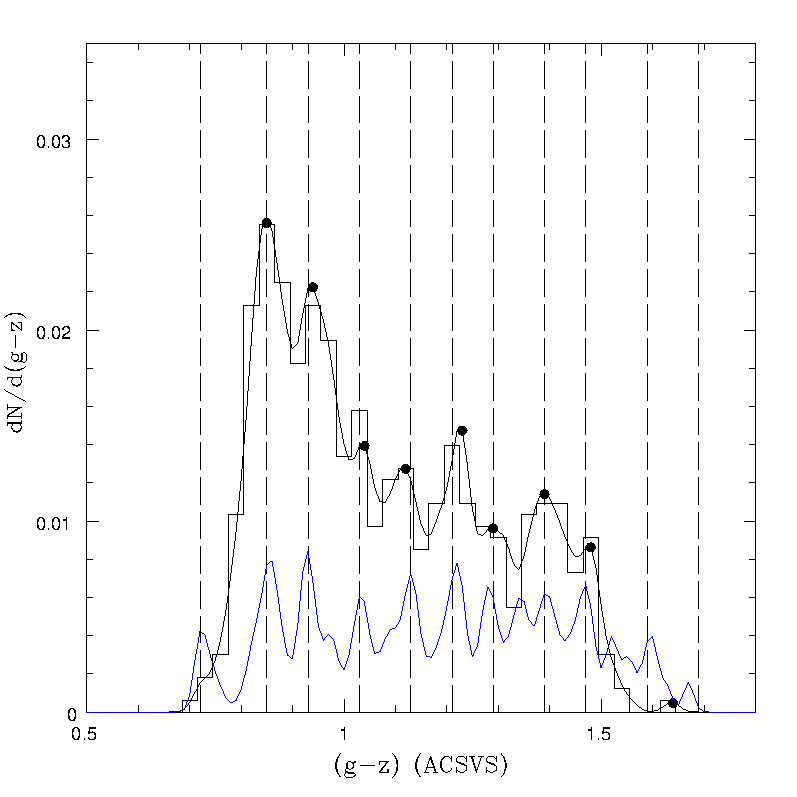}
    \caption{$(g-z)$ colour distribution corresponding to a composite sample of 543 clusters in seven galaxies with $M_{g}$ from -19.7 to -19.2. The blue line is the (arbitrarily) scaled down frequency curve shown in Fig. 1. Dashed lines correspond to the $TVP$ colours.
}
    \label{fig:fig3}
\end{figure}
\begin{figure}
\vspace{-0.5cm}
	\includegraphics[width=\columnwidth]{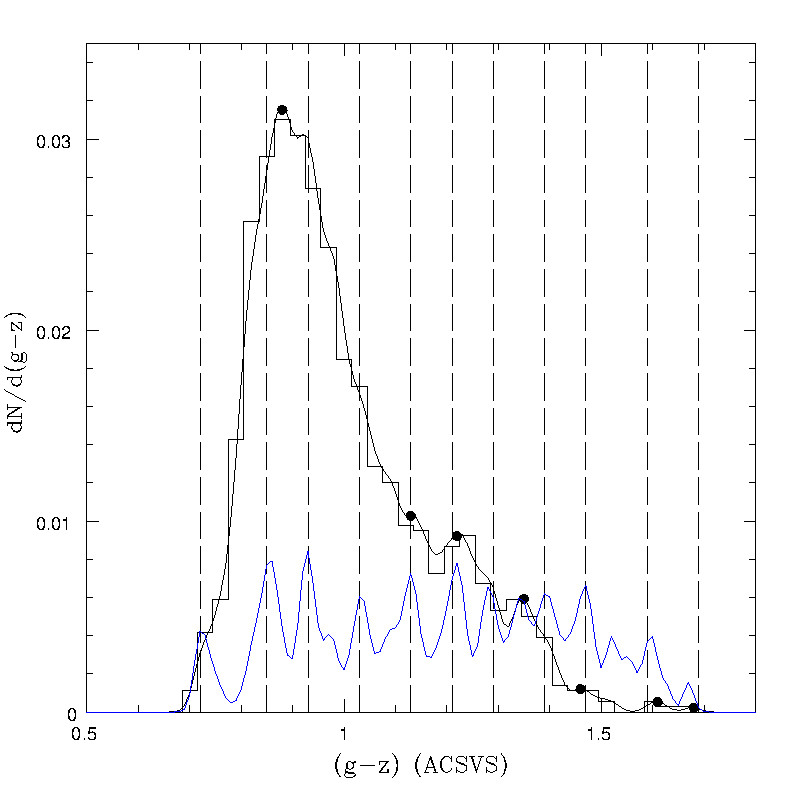}
    \caption{$(g-z)$ colour distribution corresponding to a composite sample of 1181 clusters in fifty eight Virgo galaxies with $M_{g}$ from -19.2 to -16.0. The blue line is the (arbitrarily) scaled down frequency curve shown in Fig. 1. Dashed lines correspond to the $TVP$ colours.  
}
    \label{fig:fig4}
\end{figure}
 The results presented in $F2017$ were based on the analysis of $(g-z)$ colours (ACSVS system) from \citet{Jordan2009}. Composite $GC$ samples, within a galactocentric radius ($R_{gal}$) of 110\arcsec, were taken from different galaxy groups contained in a moving sampling window, defined in the absolute magnitude $M_{g}$ vs. $(g-z)$ colour plane. 
 
  Each of those $GC$ colour samples was smoothed with a Gaussian kernel (see below) and colour peaks were identified. The statistics of these colours, in turn, define the relative frequency of appearance of each of these peaks in different samples. In what follows, these components are identified with their $(g-z)$ colours (either in the $ACSVS$ or $SDSS$ photometric systems) given within brackets.

 The size of the smoothing kernel requires a proper tuning, since a wide kernel will erase eventual colour features while a narrow one, will lead to oversampling, and spurious colour peaks.
 
 In order to have a quantitative estimate of the proper kernel size, we used $Monte-Carlo$ models. These models include Gaussian integrated luminosity functions and different $GC$ colour distributions with bimodal or multi peaked patterns. 
 
 Simulated photometric errors were added (using the $rms$ values presented in different photometric works) before proceeding to the smoothing and colour pattern analysis. For each model $GC$, the photometric errors were assigned adopting those of the nearest observed cluster neighbour in a $g$~vs.~$(g-z)$ magnitude-colour diagram. 
 
 In the case of the $ACSVS$~$(g-z)$ data from \citet{Jordan2009}, stable features are detected with a 0.035 mag~$FWHM$ Gaussian kernel ($i.e$~$\sigma_{(g-z)}=$0.015~mag). Throughout this paper, kernel sizes are given in terms of $FWHMs$.
 
  Numerical models are also useful to estimate the probability of generating a $TVP$-like $GCCD$, just as a result of sampling noise on a bimodal $GCCD$ characterized by two (blue and red) Gaussian colour distributions, and photometric errors comparable to those given by \citet{Jordan2009}.
 
 After $10^3$ realizations, only five percent of the  model outputs, aimed at reproducing multi-peaked $GCCDs$, show similarities with the $TVP$. This similarity is quantitatively defined by means of the $rms$ of the colour differences between a given model colour peak and the colour of the nearest peak in the $TVP$. 
 
 The same models were used to estimate the effectiveness of the $RSS$ approach in assessing the presence of colour peaks in a given $GCCD$. The results indicate that for a population in the order of $10^{3}$~$GC$ candidates, a sub-sampling size of about ten percent of the total, leads to the recovery of the input patterns after a few hundred sampling cycles.
 
  It must be stressed that the $RSS$ technique, $\it{''per~se''}$, is not able to discriminate between a pattern with a physical entity or, eventually, one arising in statistical fluctuations connected with stochastic processes. However, these last type of fluctuations are fragile to sub-sampling, in contrast with the characteristics of the $TVP$ features, that survive to different sampling criteria as discussed in $F2017$. 
   
  Fig.~\ref{fig:fig1} shows the $RSS$ output corresponding to 2235 $GCs$ in seventy one Virgo galaxies with $M_{g}$ between -20.2 and -16.0, adopting a sub-sampling size of 10 percent, and 5$\times$$10^{2}$ random sampling cycles, that produced 5344 colour peaks. For this sample we set a limiting magnitude $g=$24.0 (similar to that in the $OA2016$ work). The $(g-z)$ colour peak statistics is given in the form of a discrete histogram (bin size: 0.02 mag) and also after smoothing with a narrow 0.023 mag Gaussian kernel.  
   
  Dashed lines in Fig.~\ref{fig:fig1} correspond to the revised $TVP$ colours adopted in what follows, namely, $(g-z)_{ACSVS}$: 0.72; 0.85; 0.93; 1.03; 1.13; 1.21; 1.29; 1.39; 1.47; and 1.60. These colours are independent of galaxy sampling windows, and are preferred to those presented in $F2017$, shown as black dots in the upper part of the diagram.
  
  A comparison between the revised $TVP$ colours and those of the previous determination, shows small differences with maximum deviations of $\approx$ 0.02 mag. The open dot in Fig.~\ref{fig:fig1} identifies a peak at $(g-z)=$ 1.35, not detected in $F2017$. This peak mostly appears in $GC$ candidates with galactocentric radii between 30\arcsec~and 60\arcsec~associated to $NGC~4442$ ($VCC$ 1062). 
   
  The [1.72] peak, reported in $F2017$ and not present in this figure, is the result of field contamination (as seen in the following Fig.~\ref{fig:fig5} and Fig.~\ref{fig:fig6}). However, a peak at [1.69] seems present in the inner regions of $NGC~4486$ (see Section 5).
   
   The $GCCD$ for all the clusters in the sample are displayed in Fig.~\ref{fig:fig2}, Fig.~\ref{fig:fig3} and Fig.~\ref{fig:fig4}.  These distributions, and similar ones throughout the paper, are normalized by the total number of objects in each sample.
   
   The first two diagrams include half of the total $GC$ sample associated to the brightest 13 galaxies (1054 objects), split in two sub-samples (in six and seven galaxies) with approximately the same number of objects. Black dots correspond to the colour peaks found on the $GCCD$ after smoothing with a 0.035 mag Gaussian Kernel. The main features of the $TVP$ colours are common to both  diagrams.
   
   In turn, Fig.~\ref{fig:fig4} includes 1181 $GC$ candidates in 58 galaxies with $M_{g}$ in the range from -19.2 to -16.0. This sample, shows a dominant blue peak [at $(g-z)=$0.87], in contrast with the double peak ([0.85];[0.93]) detectable in the brighter galaxies    
   
\section{Reference colours frame and globular cluster candidates from photometric data}
\label{section3}  
  A problem for the comparison of the $GCCDs$ in different galaxies arises in the various photometric systems, photometric zero points, interstellar reddening corrections, and also in the criteria to define $GC$ candidates, adopted in different works. These differences may conspire against the identification of eventual common features.
   
  This work adopts the $SDSS-ugriz$ multicolour relations defined by Virgo galaxies brighter than $g=$14.0 as presented by \citet{Chen2010}. The rationale behind adopting the integrated colours of early Virgo galaxies as a reference frame, is that these systems are composed by a number of $GCs-like$ old stellar sub-populations. The similarity between the slopes of the colour-colour relations of $GCs$ and galaxies was noted, for example, in \citet{Forte2014}.  
  
 The definition of the forty five colour-colour relations that can be obtained from $ugriz$ magnitudes, the transformation of magnitudes from the $NGVS$ to the $SDSS$ system, and the identification of $GC$ candidates through pseudo-continuum fits, are described in the Appendix. In particular, the relation between the $(g-z)$ colours in both photometric systems results:
\begin{eqnarray}
\label{uno}
 (g-z)_{SDSS} = 1.090(g-z)_{NGVS} + 0.05
\end{eqnarray}
 with uncertainties of $\pm$~0.02 and $\pm$~0.015 for the slope and zero point terms, respectively.
  \section{Globular cluster candidates in NGC 4486 from NGVS data}
\label{section4}
 The present work is restricted to an area with a galactocentric radius ($R_{gal}$) range from 0.5\arcmin~to~60\arcmin, $i.e.$, not including the innermost core of $NGC~4486$, where the completeness of the $OA2016$ work is affected by the galaxy halo brightness. The studied region has a total areal coverage of 85 percent and avoids the ''Markarian's Chain'' galaxies, located towards the $NW$~of~$NGC~4486$.
 
 In what follows, and from the \citet{Schlafly2011} maps, we adopt a mean colour excess $E_{(B-V)}=$ 0.024 and then, interstellar extinction values of 0.123; 0.087; 0.065; 0.051 and 0.034 mag, for the $u,g,r,i,z$ bands, respectively.
 
 The results based on $NGVS$ data presented by \citet{Munoz2014} and \citet{Powalka2016}, show the complexity of disentangling $GC$ populations from field objects based purely on photometric data. These results also indicate that, using single colour photometry and adopting  $''reasonable''$ colour ranges to identify $GCs$, will not avoid a significant fraction of field contaminants. The presence of these field objects, in turn, could mask the eventual detection of subtle colour patterns.
 
  As an attempt to identify $GC$ candidates in this region, we assume that the empirical relations given in the Appendix, define the locus of $GCs$ in a multi-colour space. In this approach the status of $GC-candidate$ is earned if the residuals left by the fits of the $ugriz$ pseudo-continuums, listed in Table~\ref{table_A2}, fall within the acceptability boundaries defined by the $\alpha=$0.35~$\pm$0.05, and~$a=$0.03~$\pm$0.005 parameters (see the Appendix), and have a probability $P$~$>$ 0.5 of being either a genuine blue or red cluster according to $OA2016$.
  
\begin{figure}
\vspace{-0.5cm}
	\includegraphics[width=\columnwidth]{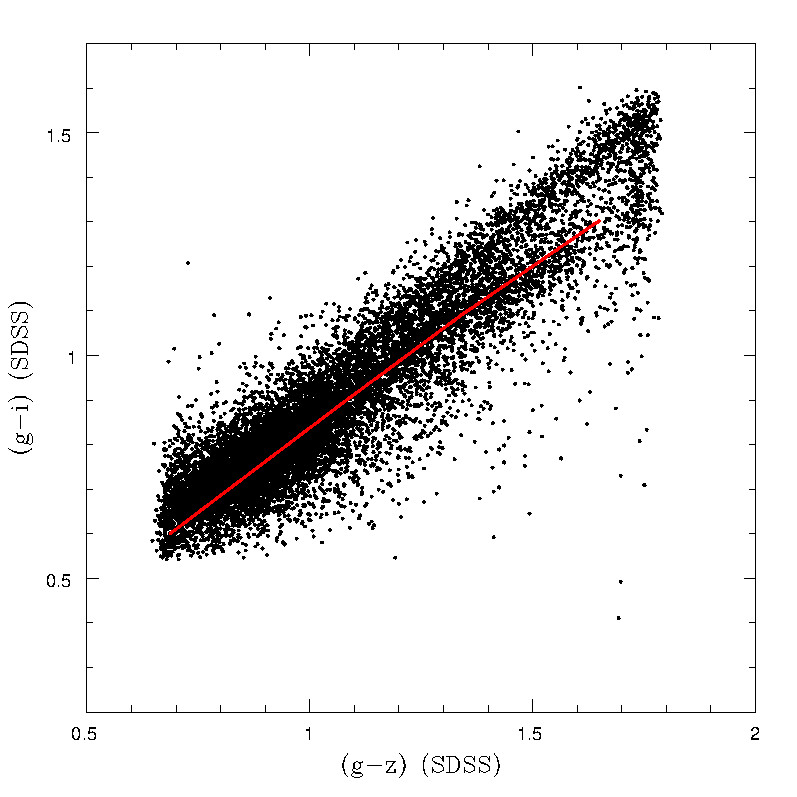}
    \caption{$(g-i)$ vs. $(g-z)$ colours diagram corresponding to 13271 objects within 60\arcmin~from the centre of $NGC~4486$. The red line indicates the relation for globular clusters derived from Table~\ref{table_A2}.
}
    \label{fig:fig5}
\end{figure}

\begin{figure}
\vspace{-0.5cm}
	\includegraphics[width=\columnwidth]{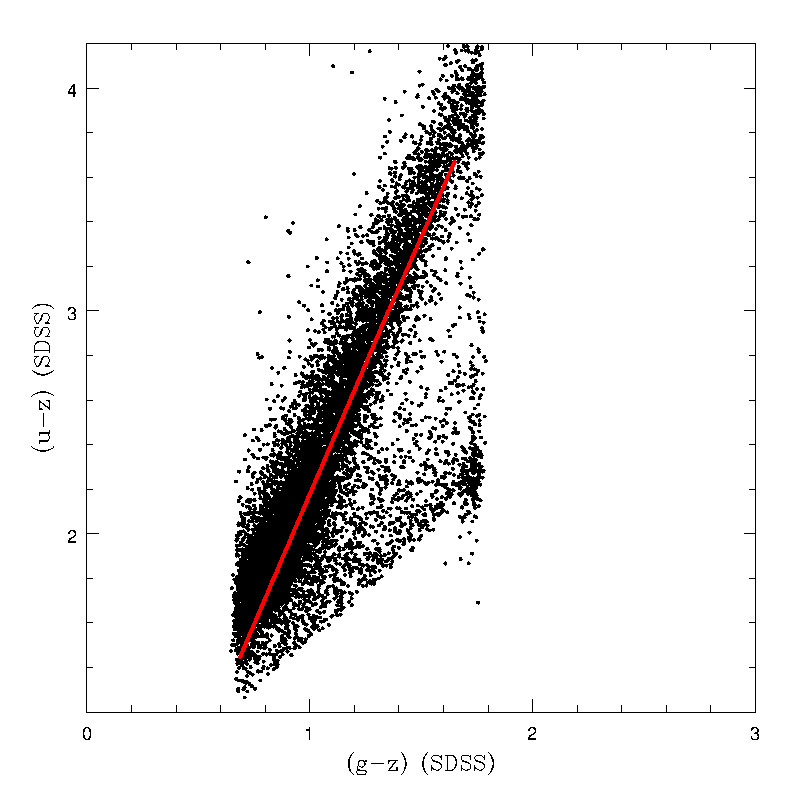}
    \caption{$(u-z)$ vs. $(g-z)$ colours  diagram corresonding to 13271 objects within 60\arcmin~from the centre of $NGC~4486$. The red line indicates the relation for globular clusters derived from Table~\ref{table_A2}.
}
    \label{fig:fig6}
\end{figure}

\begin{figure}
\vspace{-0.5cm}
	\includegraphics[width=\columnwidth]{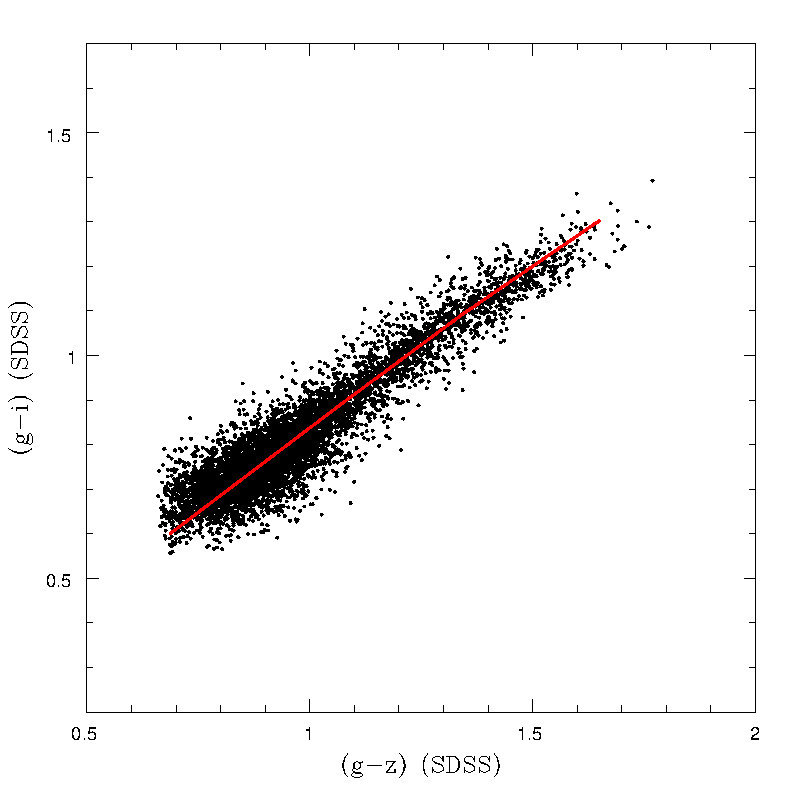}
    \caption{$(g-i)$ vs. $(g-z)$ colours  diagram for 6306 objects considered as $GC$ candidates within 60\arcmin~from the centre of $NGC~4486$. The red line indicates the colour-colour relation for globular clusters derived from Table~\ref{table_A2}.
}
    \label{fig:fig7}
\end{figure}

\begin{figure}
\vspace{-0.5cm}
	\includegraphics[width=\columnwidth]{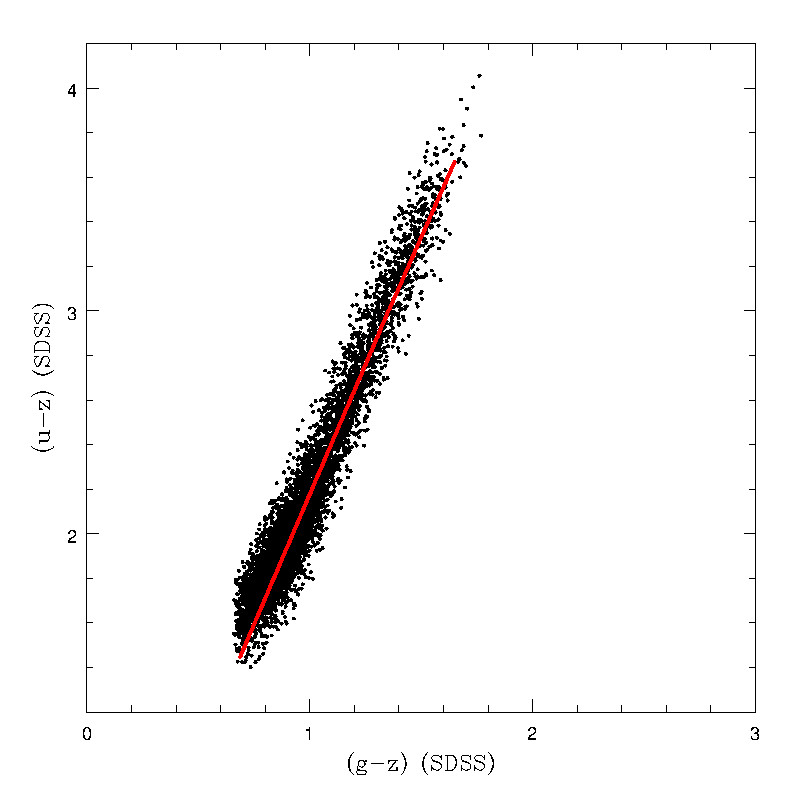}
    \caption{$(u-z)$ vs. $(g-z)$ colours  diagram for 6306 objects considered as $GC$ candidates within 60\arcmin~from the centre of $NGC~4486$. The red line indicates the relation for globular clusters derived from Table~\ref{table_A2}.
}
    \label{fig:fig8}
\end{figure}

  The $(g-i)$ vs. $(g-z)$ and $(u-z)$ vs. $(g-z)$ diagrams for all the 13271 objects brighter than $g=$24.0, within $R_{gal}=$60\arcmin~are presented in Fig.~\ref{fig:fig5} and Fig.~\ref{fig:fig6}. In turn, the same diagrams for 6306 objects considered as $GC$ candidates are displayed in Fig.~\ref{fig:fig7} and Fig.~\ref{fig:fig8}, respectively.
\begin{figure}
	\includegraphics[width=\columnwidth]{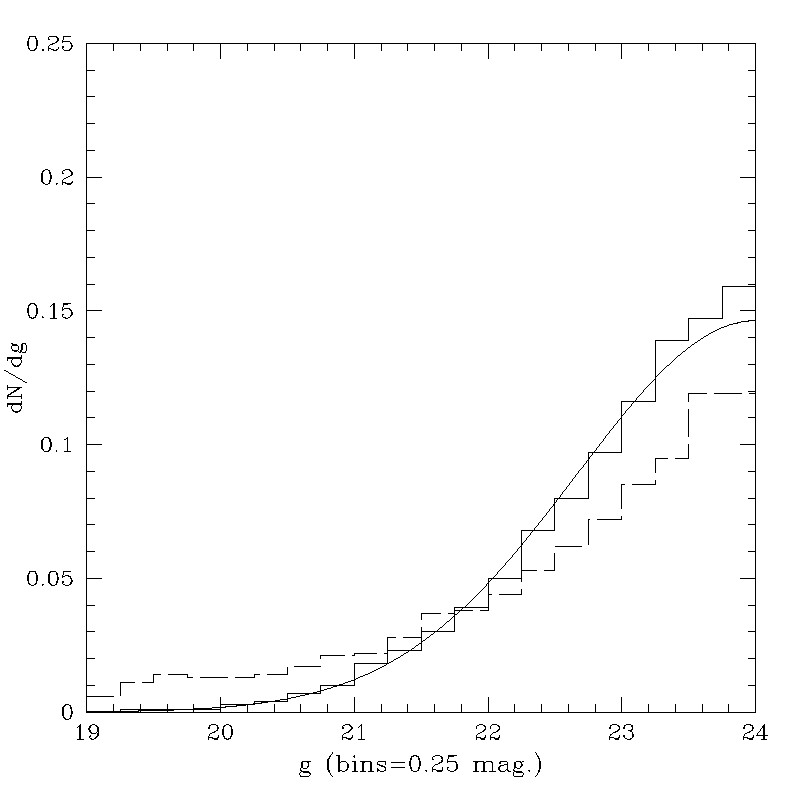}
\vspace{-0.5cm}
    \caption{Integrated GCs luminosity function for 6306 $GC$ candidates within $R_{gal}=$60\arcmin~in $NGC~4486$, . The solid curve is a reference Gaussian with $\sigma_{g}=$1.35 mag. and a turn-over at $g=$24.0. The dashed histogram corresponds to 6817 objects considered as field interlopers.
}
    \label{fig:fig9}
\end{figure}
\begin{figure}
	\includegraphics[width=\columnwidth]{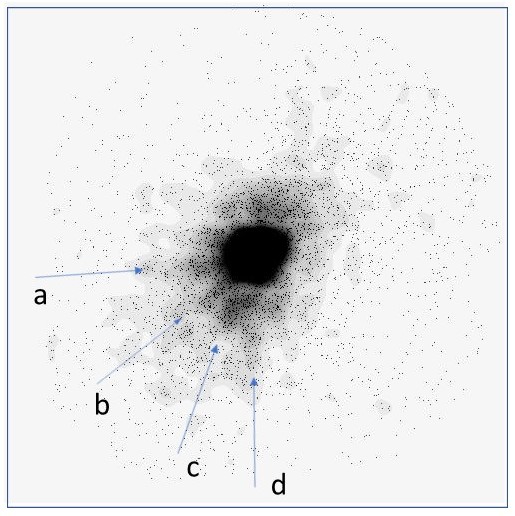}
\vspace{-0.5cm}
    \caption{Projected spatial distribution of 6306 $GC$ candidates within $R_{gal}=$60\arcmin~in $NGC~4486$. The stream-like structures are identified with labeled arrows. The horizontal size of the image is 120\arcmin. North is up; East, to the left.
}
    \label{fig:fig10}
\end{figure}
\begin{figure}
	\includegraphics[width=\columnwidth]{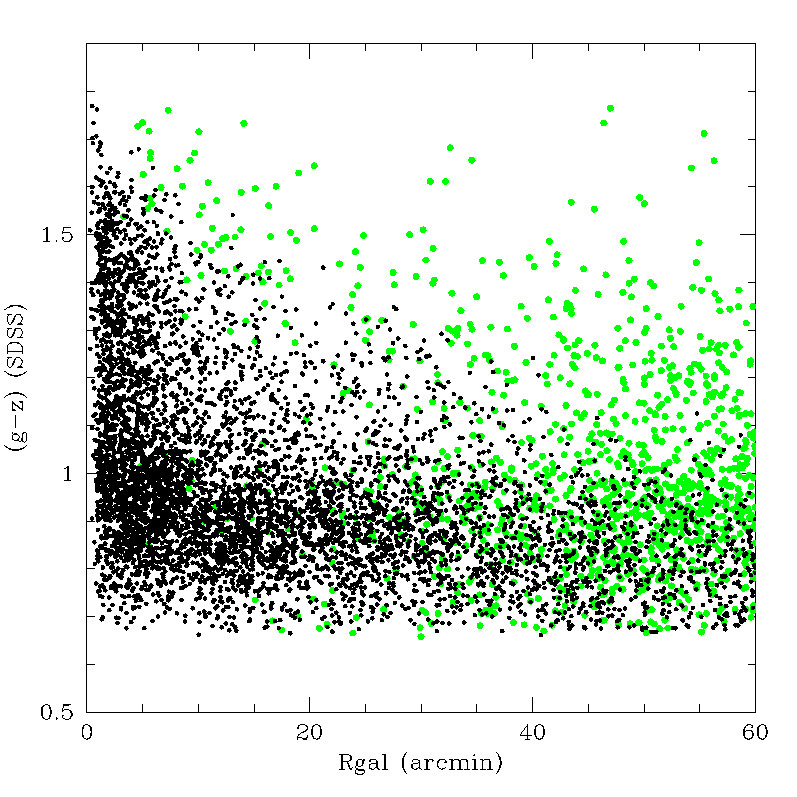}
\vspace{-0.5cm}
    \caption{$(g-z)$ colours as a function of galactocentric radius for 6306 $GC$ candidates within $R_{gal}=$60\arcmin~in $NGC~4486$. Green dots correspond to 1170 objects with a probability $P$~$<$ 0.5 (see text).
}
    \label{fig:fig11}
\end{figure}
  
 The integrated $GCs$ luminosity function for all the $GC$ candidates within $R_{gal}=$60\arcmin~is displayed in Fig.~\ref{fig:fig9}. This radius corresponds to 290~$Kpc$, adopting a distance modulus $V-M_{V}=$31.1, derived by \citet{Tonry2001}.
 
 As a reference, Fig.~\ref{fig:fig9} includes a Gaussian function characterized by a turn over magnitude $g=$24.0 and a dispersion $\sigma_{g}=$1.35 mag. These parameters are representative for $GCs$ associated with giant galaxies in Virgo \citep[e.g.][]{Villegas2010}.
  
  The $g$ magnitude distribution for objects considered as field interlopers ($MW$ stars and unresolved extragalactic objects), discussed in detail in $OA2016$, is shown as dashed lines. 
  
  The projected position of the $GC$ candidates on the sky is seen in Fig.~\ref{fig:fig10}. This diagram combines the position of these globulars with an smoothed image (adopting a 1.75\arcmin~bi-dimensional Gaussian kernel). The  combined image exhibits several azimuthal inhomogeneities. The presence of spatial structures that seem the result of recent or ongoing galaxy mergers, has been noted in the extensive analyisis of $GCs$ in the $Virgo~cluster$ by \citet{Durrell2014}.
  
  Four of the most prominent radial stream-like structures in Fig.~\ref{fig:fig10} are labeled as $\bf{a}$, $\bf{b}$, $\bf{c}$, and $\bf{d}$. In particular, the feature identified as $\bf{a}$ has been discussed by \citet{Romanowsky2012}, using radial velocity data presented in \citet{Strader2011} ($ST2011$ in what follows). Their analysis indicates that in fact, this feature is a tidal stream delineated by $GCs$ over a range of $\approx$~35\arcmin. 
  
  The $(g-z)$ vs $R_{gal}$ values for $GC$ candidates within $R_{gal}=$60\arcmin,~is displayed in Fig.~\ref{fig:fig11}. This figure also includes 1170 objects (green dots) that fall inside the photometric acceptability boundaries, but have $P$~$<$ 0.50 and were considered as field interlopers. These objects show systematic deviations from the $GC$ colours locus, and no concentration towards the galaxy centre.
   
  Fig.~\ref{fig:fig11} is comparable, for example, to that presented by \citet{Ko2022}, using the $(g-i)_{NGVS}$ colours, on a larger galactocentric range. In a $vertical$ reading of their Fig. 2, these authors detect a small colour gradient for red $GCs$ and a steeper one for blue $GCs$. An alternative view of this kind of diagrams is given in Section 6.
  
\begin{figure}
	\includegraphics[width=\columnwidth]{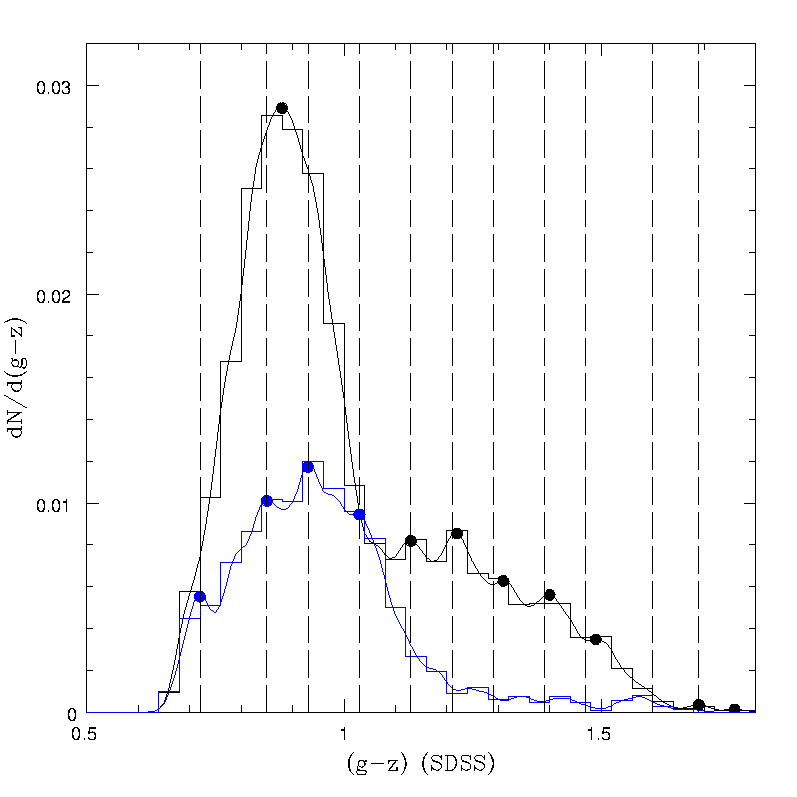}
\vspace{-0.5cm}
    \caption{$(g-z)$ colour distribution for 4474 $GC$ candidates with $P$~$>$ 0.90 (black line) and for 1832 $GC$ candidates with 
    $P$ from 0.50 to 0.90 (blue line), within $R_{gal}=$60\arcmin~in $NGC~4486$. Dashed lines correspond to the 
    $TVP$~$(g-z)$ colours in the $SDSS$ photometric system. 
}
    \label{fig:fig12}
\end{figure}

  A first hint about the presence of the $TVP$ within a galactocentric radius of~60\arcmin~is provided by Fig.~\ref{fig:fig12}. This diagram corresponds to the $(g-z)$ $GCCD$ of a sub-sample of 4474 $GC$ candidates, with $P$~$>$ 0.90 (black line). This distribution shows a prominent blue peak at $(g-z)=$ 0.87, close to that observed in the composite $GCs$ sample corresponding to galaxies fainter than  $M_{g}=$-19.0 (see Fig.~\ref{fig:fig4}), and other five redder peaks at the expected positions of the $TVP$ components.
  
  Fig.~\ref{fig:fig12} also includes the $GCCD$ for 1832 $GC$ candidates with $P$ from 0.50 to 0.90 (blue line). This sample shows barely detectable colour peaks but, again coincident with those of the $TVP$.
      
\section{Colour pattern in the GCCD of globular clusters associated to NGC 4486}
\label{section5}
 The analysis of both the $GCCDs$ and of the $RSS$ otputs as a function of galactocentric radius and position angles, finds a region between $R_{gal}=$5\arcmin~and 9\arcmin~and $P.A.$ from 270\degr~to 345\degr,~where the colour pattern is not clearly defined. This region includes 268 objects that, both on statistical and photometric terms, can be considered as genuine $GC$ candidates. These objects constitute $\approx$ 4 percent of the total sample and were removed for the following discussion. This leaves a total of 6038 $GC$ candidates for the analysis of the colour pattern.

 The $RSS$ routine was run on this last sample (performing 5$\times$$10^{2}$ random sampling cycles), on both the $OA2016$-$NGVS$ $(g-z)$ colours, and on their transformed version to the $SDSS$ system. The correlation of the $TVP$ peaks found in the $SDSS$ system, as well as those in the $ACSVS$ system (discussed in Section 2), against the $NGVS$ peaks, are presented in Fig.~\ref{fig:a5}. This diagram indicates that there are no significant differences between the $TVP$ $(g-z)$ colours in the $ACSVS$ (filled dots) and in the $SDSS$ (open dots) systems.
 
 Fig.~\ref{fig:a5} also includes the $(g-z)$ colour relation derived in Section 3 (blue line), which is in very good agreement with that presented by \citet{Wu2022} (red line). Their relation was shifted by $+0.015$ mag in ordinates, indicating the existence of a small zero point difference with this work.

 The $TVP$  $(g-i)$, $(g-z)$, and $(u-z)$ colours in the $SDSS$ system assuming, from the previous result, that the $(g-z)$ colours are coincident with those in the $ACSVS$ system, are listed in Table~\ref{table_A3}. 
 
\begin{figure}
	\includegraphics[width=\columnwidth]{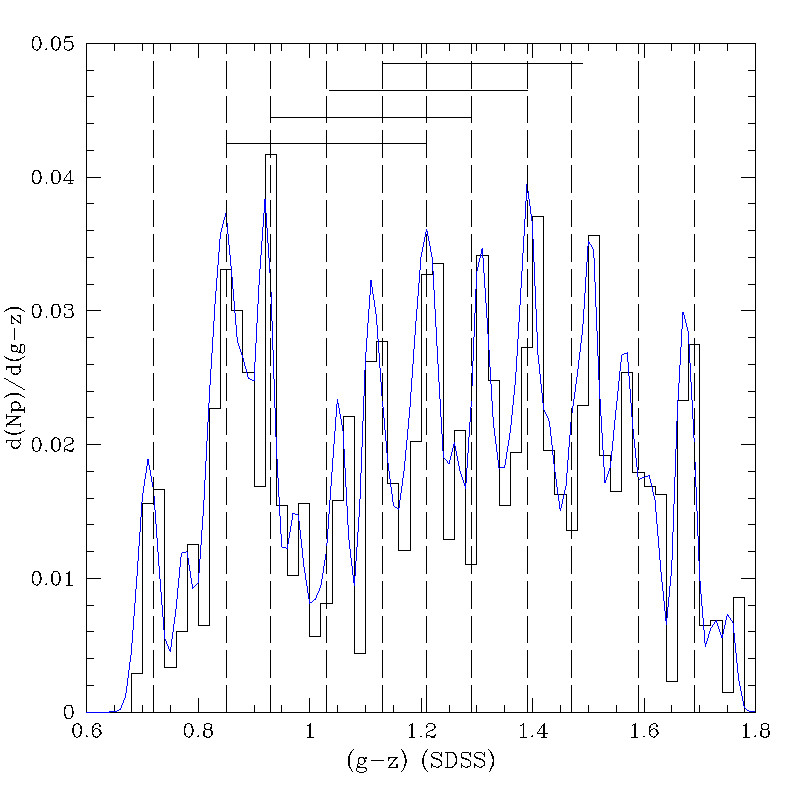}
\vspace{-0.5cm}
    \caption{Relative frequency of $(g-z)$ colour peaks found by the $RSS$ routine on a sample of 6038 $GC$ candidates within $R_{gal}=$60\arcmin in $NGC~4486$. Vertical dashed lines indicate the $TVP$ colours discussed in Section 2. Horizontal lines (linking given blue and red peaks) have a length of 0.36 mag (see Section 5).
}
    \label{fig:fig13}
\end{figure}

 The output from the $RSS$ routine, corresponding to 6038 $GC$ candidates within $R_{gal}=$60\arcmin~and using the same smoothing kernel (0.035 mag) adopted in Section 2, is presented in Fig.~\ref{fig:fig13}. This diagram includes 4799 colour peaks, after 5$\times$$10^{2}$ sampling cycles. A comparison with the $TVP$ colours listed in Table 3 yields $\sigma_{(g-z)}=$ 0.016, and a maximum deviation of +0.03 mag from that pattern, occurs for the peak at $(g-z)=$1.50. 
 
  By contrast, none of these features are detectable in the $(g-z)$ colour distribution of 6817 objects considered as field interlopers (see Fig.~\ref{fig:a6}).
 
 Both the nature of the peak at [1.69], in Fig.~\ref{fig:fig1}, and the meaning of the horizontal lines in Fig.~\ref{fig:fig13}, linking given blue and red colour peaks separated by 0.36 mag intervals, are discussed in Subsection 5.2.   
  
\subsection{Globular clusters colour pattern in the inner 5\arcmin~region of NGC~4486}
\label{ss_5a}
 The $RSS$ analysis of $GC$ candidates in different galactocentric regions, indicates that most of the pattern seen in Fig.~\ref{fig:fig13} arises close to the centre of $NGC~4486$.
 
\begin{figure}
	\includegraphics[width=\columnwidth]{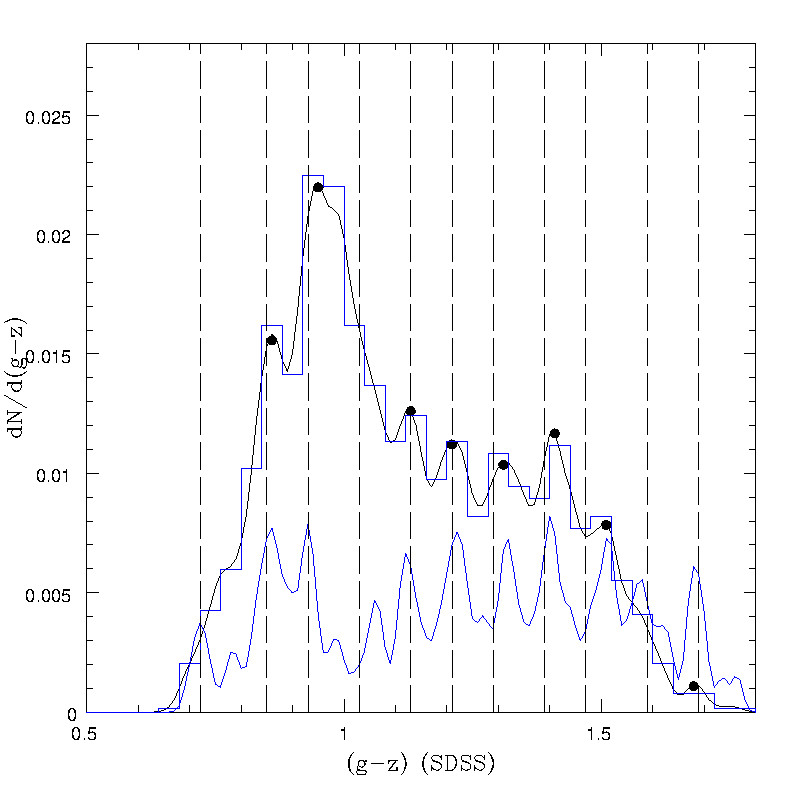}
\vspace{-0.5cm}
     \caption{$(g-z)$ colour distribution for 1591 $GC$ candidates within a galactocentric radius of~5\arcmin~in $NGC~4486$. Dashed lines correspond to the $TVP$ colours. The lower blue curve is the (arbitrarily scaled down)  colour peak frequency displayed in Fig.~\ref{fig:fig13}.
}
    \label{fig:fig14}
\end{figure}

The $(g-z)$~$GCCD$ for 1591 $GC$ candidates within $R_{gal}=$5\arcmin,~displayed in Fig.~\ref{fig:fig14}, shows the presence of seven peaks coincident with the $TVP$ colours ($\sigma_{(g-z)}=$ 0.012 mag). The arbitrarily scaled down $RSS$ output displayed in Fig.~\ref{fig:fig13}, is also included in this diagram (lower blue curve).

\begin{figure}
	\includegraphics[width=\columnwidth]{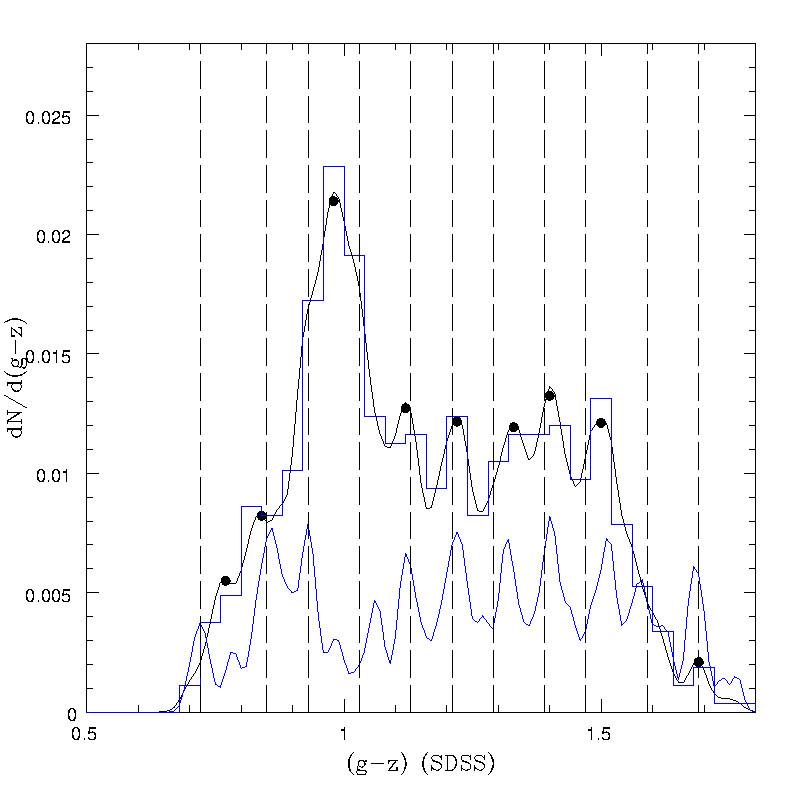}
\vspace{-0.5cm}
    \caption{$(g-z)$ colour distribution for 667 $GC$ candidates within a galactocentric radius of 2.5\arcmin~in $NGC~4486$. Dashed lines correspond to the $TVP$ colours. The lower blue curve is the (arbitrarily scaled down)  colour peak frequency displayed in Fig.~\ref{fig:fig13}
}
    \label{fig:fig15}
\end{figure}

 Restricting the analysis to  $R_{gal}<$ 2.5\arcmin~, reveals the presence of a single blue peak (with an extended tail towards the blue) as seen in Fig.~\ref{fig:fig15}. A similar blue peak, and colour peak structure, is found by \citet{Bellini2015} in their $HST$ photometric survey, including the ultraviolet filter $F275W$, in the same area (see Fig. 6-c1 in that work).

\begin{figure}
	\includegraphics[width=\columnwidth]{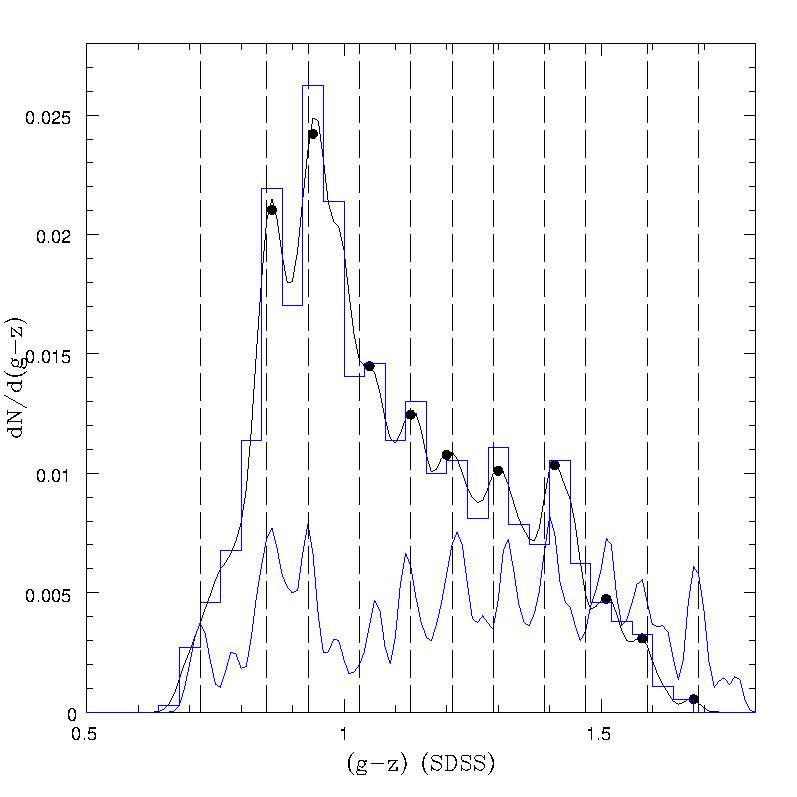}
\vspace{-0.5cm}
    \caption{$(g-z)$ colour distribution for 924 $GC$ candidates within a galactocentric radius range from 2.5\arcmin~to~5\arcmin~in $NGC~4486$. Dashed lines correspond to the $TVP$ colours. The lower blue curve is the (arbitrarily scaled down)  colour peak frequency displayed in Fig.13. 
}
    \label{fig:fig16}
\end{figure}

In turn, Fig.~\ref{fig:fig16} corresponds to 924 $GC$ candidates in the $R_{gal}$ range from 2.5\arcmin~to~5\arcmin. This figure shows a decrease in the number of red clusters, a prominent double blue peak structure ([0.87] and [0.95]), comparable to those seen in Fig.~\ref{fig:fig2} and Fig.~\ref{fig:fig3}, as well as other five redder colour peaks, consistent with the $TVP$ colours ($\sigma_{(g-z)}=$~0.011 mag). 

  A further scrutiny about the presence of the $TVP$ within $R_{gal}=$5\arcmin, can be performed using $GCs$ with measured radial velocities and $(g-i)$ colours as presented by $ST2011$. In this last work, the original $(g-i)$ colours were taken from \citet{Harris2009} and were transformed to the $SDSS$ photometric system. The sample includes 340 $GCs$, as well as 39 objects with half light radii $>$ 5~$Pc$ (classified as "extended" or $UCD$ objects), in a $g$ magnitude range from 19.0 to 23.5 and radial velocities from 500 {\,km\,s$^{-1}$} to 2500 {\,km\,s$^{-1}$}.	
 
\begin{figure}
	\includegraphics[width=\columnwidth]{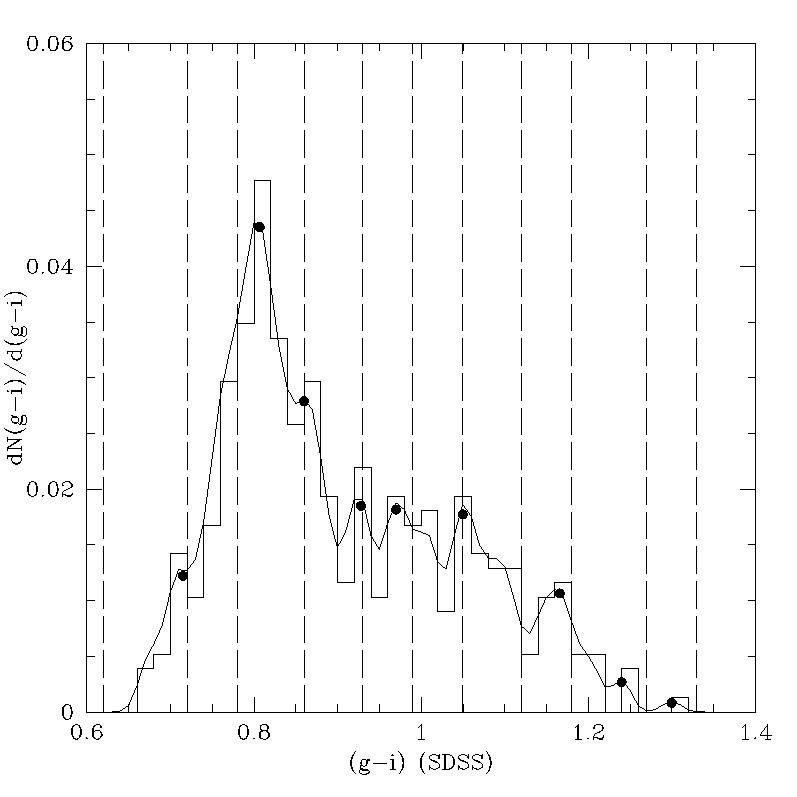}
\vspace{-0.5cm}
    \caption{$(g-i)$ colour distribution for 379 $GCs$ with measured radial velocities (from Strader et al. 2011), within a galactocentric radius of~5\arcmin~in $NGC~4486$. Dashed lines correspond to the $(g-i)$ $TVP$ colours as given in Table~\ref{table_A3}.
}
    \label{fig:fig17}
\end{figure}

 The $(g-i)$ colour distribution for these objects is displayed in Fig.~\ref{fig:fig17}, where dashed lines represent the $TVP$~$(g-i)$ colours listed in Table~\ref{table_A3}. Taking into account that the sensibility of $(g-i)$ to metallicity is about seventy percent of that of the $(g-z)$ colour, the smoothing Gaussian kernel was set to 0.026 mag. 

  Fig.~\ref{fig:fig17} preserves most of the characteristics of the $GCCD$ seen in Fig.~\ref{fig:fig14}, $i.e.$, seven peaks consistent with the $TVP$ colours, and an overall $\sigma_{(g-i)}=$0.017 mag. We note that the feature at $(g-i)$~$\approx$~1.0 [or $(g-z)=$1.22] seems broader than the other red peaks, as also suggested by the two colour analysis presented below. 
 
\begin{figure}
	\includegraphics[width=\columnwidth]{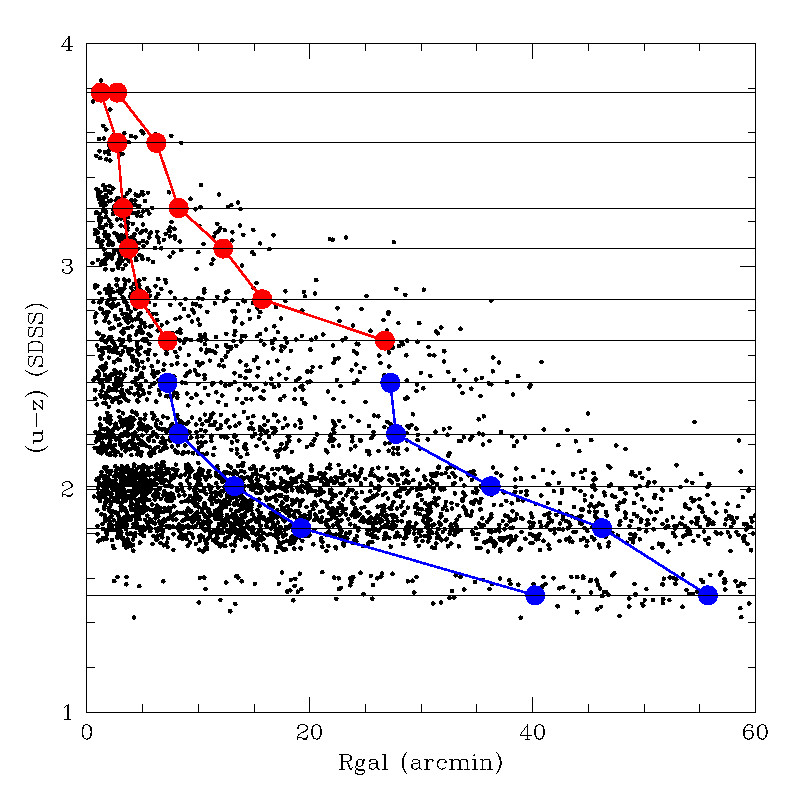}
\vspace{-0.5cm}
    \caption{$(u-z)$ colours vs $R_{gal}$ for 4008 $GC$ candidates within sampling eleven volumes centered on the $TVP$ components in a $3-D$ colour space. Blue and red lines correspond to the $R_{med}$ and the $R_{90}$ radii of blue and red $GCs$ (see text). Horizontal lines are the $(u-z)$~$TVP$ colours listed in Table~\ref{table_A3}.
}
    \label{fig:fig18} 
\end{figure}
 \begin{figure}
	\includegraphics[width=\columnwidth]{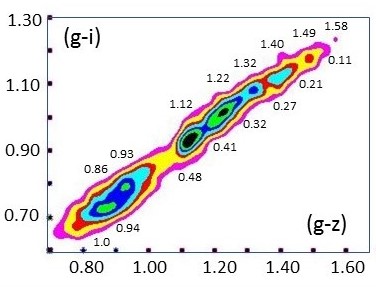}
\vspace{-0.5cm}
    \caption{Smoothed $(g-i)$ vs. $(g-z)$ colours (SDSS system) distribution for 5647  $GC$ candidates within a galactocentric range of 45\arcmin~ in NGC 4486. Each colour peak, compatible with the $TVP$, is identified by the $(g-z)$ colour of its maximum value (above) and relative heights normalized to the [0.86] peak value (below). 
}
    \label{fig:fig19}
\end{figure}
 \subsection{Spatial distribution of the TVP components and GC colour gradients}
\label{ss_5c}
 The determination of the $GC$ areal density profiles is usually performed in terms of $R^{1/4}$, power laws, or of S\'ersic profiles \citep{Sersic68}.
 
 However, a first approach to analyse the spatial distributions of each $TVP$ components, avoiding the election of a given dependence, can be performed  by determining their median galactocentric radius ($R_{med}$), and the radius that contains 90 percent of the total population ($R_{90}$). This is a $horizontal$ reading of the data as an alternative to the $vertical$ reading in \citet{Ko2022} (see their Fig. 2). 
 
 In order to isolate the $GCs$ associated with a given $TVP$ component, we adopted a $3-D$ colour space defined in terms of the $(g-i)$, $(g-z)$, and $(u-z)$ colour indices. This last colour index is the most affected by photometric errors but it is a good discriminant between genuine $GCs$ and field objects (see Fig.~\ref{fig:fig6} and Fig.~\ref{fig:fig8}).
 
  The mean separation of the $TVP$ components in the $(g-z)$ colour is $\approx$~0.09 mag and, taking into account the slopes of the relations between these three colours, this separation becomes 0.22 mag in the $3-D$ colour space. Then, we used volumetric sampling radii of 0.11 mag (i.e., the $3-D$ peak-to-valley separation), centered on each of the eleven $TVP$ colours listed in Table~\ref{table_A3}. 
 
 The distribution of the $(u-z)$ colours of 4008 $GC$ candidates contained in these eleven volumetric samples, as a function of galactocentric radius, is presented in Fig.~\ref{fig:fig18}, where horizontal lines indicate the position of the $(u-z)$~$TVP$ peaks listed in Table~\ref{table_A3}. The pattern component at $(u-z)=$3.76 (or [1.69]) corresponds to the reddest feature seen in Fig.~\ref{fig:fig13}.
 
  Fig.~\ref{fig:fig18} shows a continuous increase of both $R_{med}$ and $R_{90}$ as the $GC$ colours become bluer, for all the $TVP$ components. In the case of red $GCs$, the $R_{med}$ parameter increases from 1.25\arcmin~to 7.25\arcmin,~and from 7.25\arcmin~to 40\arcmin~for the blue $GCs$.
  
   A ''glitch'' in these trends, for both characteristic radii, is detected at $(u-z)$~$\approx$~2.50,  that corresponds to $(g-z)$~$\approx$1.15. This feature is persistent to different sampling criteria (e.g., to changes in the limiting magnitude, or position angle ranges). 
 
 The variation of the $R_{med}$ and $R_{90}$ radii with integrated colours, that can also be seen for $GC$ candidates in the colour ''valleys'', imply that the customary approach of characterizing blue and red $GCs$ by means of two different areal density profiles, is only a first approximation to a more complex situation.
 
 In their kinematic study of planetary nebulae and $GCs$ in $NGC~44486$, \citet{Doherty2009} derive an outer limit for the stellar halo at $R_{gal}$ $\approx$~35\arcmin. In turn, \citet{Durrell2014} find that red $GCs$, disappear at $R_{gal}=$45\arcmin. These results are consistent with Fig.~\ref{fig:fig18}. The two bluest $TVP$, components ([0.72] and [0.85]), in fact reach the outer galactocentric limit of this work at $R_{gal}=$60\arcmin, as also seen in the areal density profile for blue $GCs$ presented by those last authors.

 The smoothed version of the $(g-i)$ vs. $(g-z)$ colours relation for 5647 $GC$ candidates within $R_{gal}=$45\arcmin~is displayed in 
 Fig.~\ref{fig:fig19}. This diagram was obtained by mapping those colours on a 501$\times$501 matrix, that was transformed to a $fits$ image, and smoothed with a bi-dimensional Gaussian 0.035 mag kernel, through the $irafil$ and $gauss$ routines in $IRAF$. In this diagram the $GC$ population with $(g-z)$~$<$ 1.09 has been scaled down by a factor 0.35 in order to decrease the ratio of blue to red clusters and to allow the simultaneous display of all the detected colour peaks.
 
 This last figure shows two dominant  blue peaks, and  other six redder features, identified by the $(g-z)$ colours of their maximum values and relative heights (normalized to the peak value at $(g-z)$=0.86). A comparison of these eight features with the $TVP$ colours yields $\sigma_{(g-z)}=$ 0.018. In particular, the [1.05] peak is not seen, as it is located on the steep red wing of the [0.94] peak (although that peak is present in Fig.~\ref{fig:fig1}, Fig.~\ref{fig:fig2}, Fig.~\ref{fig:fig3}, Fig.~\ref{fig:fig13} and Fig.~\ref{fig:fig17}).
 
 The detectability of the $TVP$ over a $R_{gal}$ range of 45\arcmin, suggests null or very small galactocentric colour gradients for $\bf{each}$ of the individual pattern components (but see Section 7). This conclusion is also supported by Fig.~\ref{fig:a7} that corresponds to the smoothed $GCCD$ (Gaussian kernel: 0.035 mag) of three equal number samples ($\approx$1350 $GC$ candidates each, with $P$ $>$ 0.90 and $R_{gal}$ ranges from 0.5\arcmin~to 5\arcmin,~5\arcmin~to 14\arcmin~and 14\arcmin~to 45\arcmin.
 
 Fig.~\ref{fig:a7} shows that the colour peaks appear approximately at the same positions (with differences of, at most, 0.03 mag), regardless of the galactocentric range of the sample. The red lines and dots correspond to the innermost sample and exhibits the reddest detected peak at $(g-z)=$1.50. This peak dissappears in the intermediate sample (black lines and dots), while the red $TVP$ features become less evident in the outer sample (blue lines and dots).
 
 Regarding the individual components of the $TVP$, we note that:
 \begin{enumerate}
 \item The nature of the bluest peak ([0.72]) is not clear, as $GC$ candidates do not show a definite concentration towards the centre of the galaxy. An open question connected with these objects is if, at least a fraction of them, might belong to a family of intra-cluster $GCs$ \citep{Ko2018}.
 \item The [0.85] peak is barely present in the central region of the galaxy but becomes detectable for $R_{gal}$~$>$ 2.5\arcmin.
 \item The bluest peaks, at [0.72] and [0.85], reach and exceed, the outer galactocentric boundary in this work. 
 \item The [1.03] peak is the less evident blue feature in the $TVP$. Its detectability is difficult in the $GCCDs$, as it lies on the red side of the prominent [0.93] blue peak. However, its presence  is seen in two different $RSS$ outputs (see Fig.~\ref{fig:fig1} and Fig.~\ref{fig:fig13}), and also in the $GC$ sub-sample with radial velocities, at $(g-i)=$0.86 (see Fig.~\ref{fig:fig17}).
 \item The colour features at [1.21], [1.29], [1.39] and [1.47] appear as the most evident red peaks. In particular, the [1.21] peak seems a somewhat broad feature as seen both in Fig.~\ref{fig:fig17} and in Fig.~\ref{fig:fig19}.
 \item The reddest peak ([1.72]) found in $F2017$, appears in a region where the $TVP$ definition is not clear and a certain degree of field contamination would be expected (see Fig.~\ref{fig:fig5} and Fig.~\ref{fig:fig6}). However, the $ugriz$ fits identify $GC$ candidates clearly concentrated inside $R_{gal}=$3\arcmin~, producing a peak at [1.69], and suggesting that they are genuine clusters.
 \item Excluding the [0.72] peak that, as noted, has a dubious nature in terms of its connection to the galaxy, the $(g-z)$ colour separations between the remaining ten $TVP$ components range from 0.08 to 0.13 mag in $(g-z)$.
 
  The $"best"$ association between blue and red colour peaks, in terms of uniform colour separations, occurs for the [0.85]-[1.21], [0.93]-[1.29], [1.03]-[1.39], and [1.13]-[1.47] colour pairs, as seen in Fig.~\ref{fig:fig13}. In this case, the mean colour difference is 0.36 $\pm$~0.008 mag. This behaviour would be compatible with a coeval blue-red peak connection if the age metallicity and colour-metallicity relations were linear (or close to).
   
  In this case, the lack of blue counterparts for the reddest, much less prominent, and presumably youngest peaks, at [1.60] and [1.69], would suggest that the formation of blue $GCs$ may have ceased before the formation of these red $GCs$. 
 \end{enumerate}
  \section{NGC 5128: At the outskirts of the Virgo Super-Cluster}
\label{s6}
 $NGC~5128$, the nearest elliptical galaxy to the $MW$, is the dominant system in the $Centaurus~A~Group$. Adopting a distance of $\approx$ 4 $Mpc$  \citep{Harris2004a}, places this galaxy at $\approx$ 16 $Mpc$ from the centre of the $Virgo~Super-cluster$.
 
  The difficulties in disentangling its $GCs$ system from the heavy field contamination are clearly described in \citet{Harris2004b}, \citet{Harris2004a} and \citet{Woodley2010}.
 
 Recently, \citet{Hughes2021} have presented a comprehensive catalogue with 40502 $GC$ candidates, using data from $GAIA$ to remove field contaminants, and assign to each object a given total likelihood of being a genuine $GC$. As $GCs$ in $NGC~5128$ are about three magnitudes brighter than those of their counterparts in $NGC~4486$, the inherent photometric errors are considerably smaller than in the last case. This allows a better recovery of colour structures using indices that include $u$ magnitudes in their definition.
 
 In particular, the \citet{Hughes2021} photometric catalogue gives (source by source) reddening corrected $(u-r)$, and $(r-z)$ colours for a sample of their $GC$ candidates, taken from the $NOAO-Source~Catalog$ ($NSC-DEC$ camera) \citep{Nidever2018}.
 
 A comparison of the $NSC$ colour relations with those from Table~\ref{table_A2}, shows identical slopes, and that the transformation to the $SDSS$ system can be achieved trough zero point shifts:
\begin{eqnarray}
\label{dos}
(r-z)_{SDSS} = (r-z)_{NSC}+0.09\\
(u-r)_{SDSS} = (u-r)_{NSC}-0.06
\end{eqnarray}
 and hence:
 \begin{eqnarray}
\label{tres}
(u-z)_{SDSS} = (u-z)_{NSC}+0.03
\end{eqnarray}
with uncertainties of $\pm$0.01 mag for each of the zero point shifts.
 \begin{figure}
	\includegraphics[width=\columnwidth]{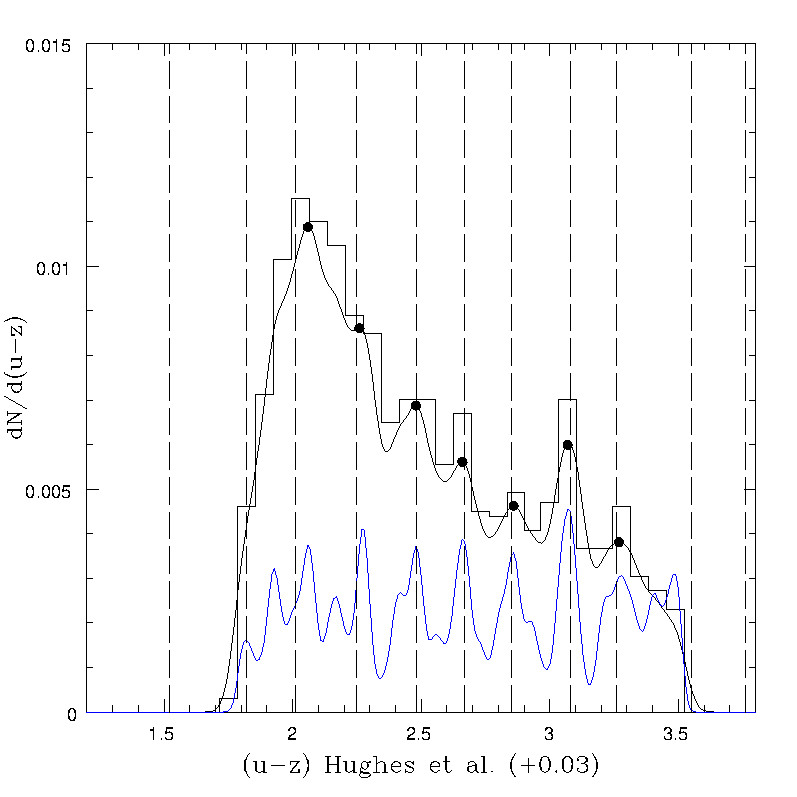}
\vspace{-0.5cm}
    \caption{$(u-z)$ colour distribution for 1480 $GC$ candidates with a likelihood $>$ 0.5, and within 2\degr~from the centre of $NGC~5128$, using data from \citet{Hughes2021}. The smoothed colour distribution was obtained adopting a 0.07 mag Gaussian kernel. Vertical dashed lines correspond to the $TVP$ pattern listed in Table~\ref{table_A3}. The blue line corresponds to the (arbitrarily scaled down) output from the $RSS$ routine on this $GC$ sample. The original $NSC$ colours have been transformed to the $SDSS$ system by adding +0.03 mag (see text).   
}
    \label{fig:fig20}
\end{figure}
 \begin{figure}
	\includegraphics[width=\columnwidth]{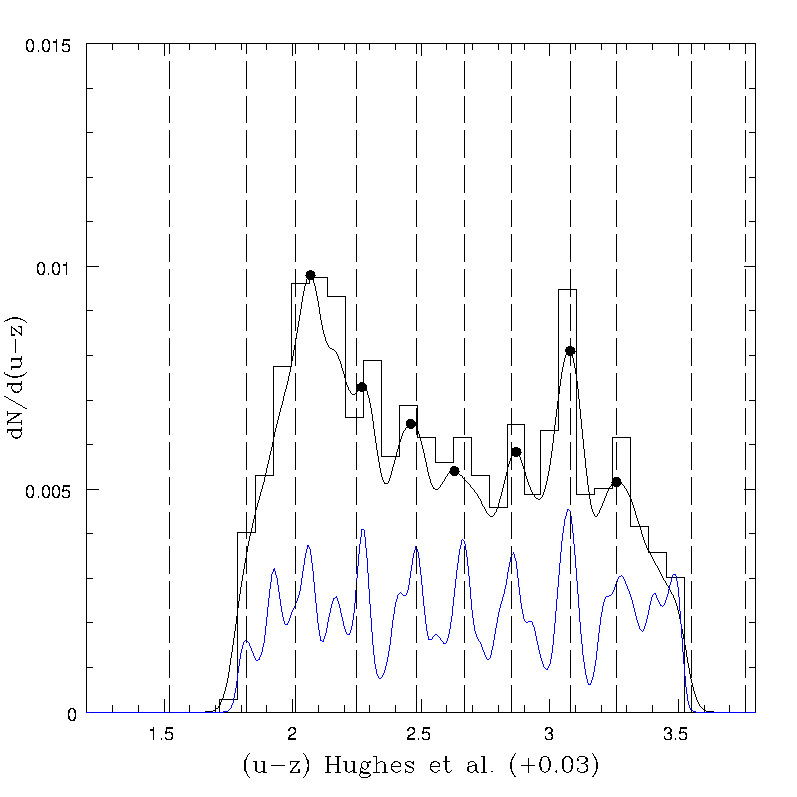}
\vspace{-0.5cm}
    \caption{$(u-z)$ colour distribution for 1080 $GC$ candidates with a likelihood $>$ 0.5, brighter than $r=$20.5, and within 2\degr~from the centre of $NGC~5128$, using data from \citet{Hughes2021}. The smoothed colour distribution was obtained adopting a 0.07 mag Gaussian kernel. Vertical dashed lines correspond to the $TVP$ pattern listed in Table~\ref{table_A3}.The blue line is the output from the $RSS$ routine run on the $GC$ sample included in Fig. 20. The original $NSC$ colours have been transformed to the $SDSS$ system by adding +0.03 mag (see text). 
}
    \label{fig:fig21}
\end{figure}
 \begin{figure}
	\includegraphics[width=\columnwidth]{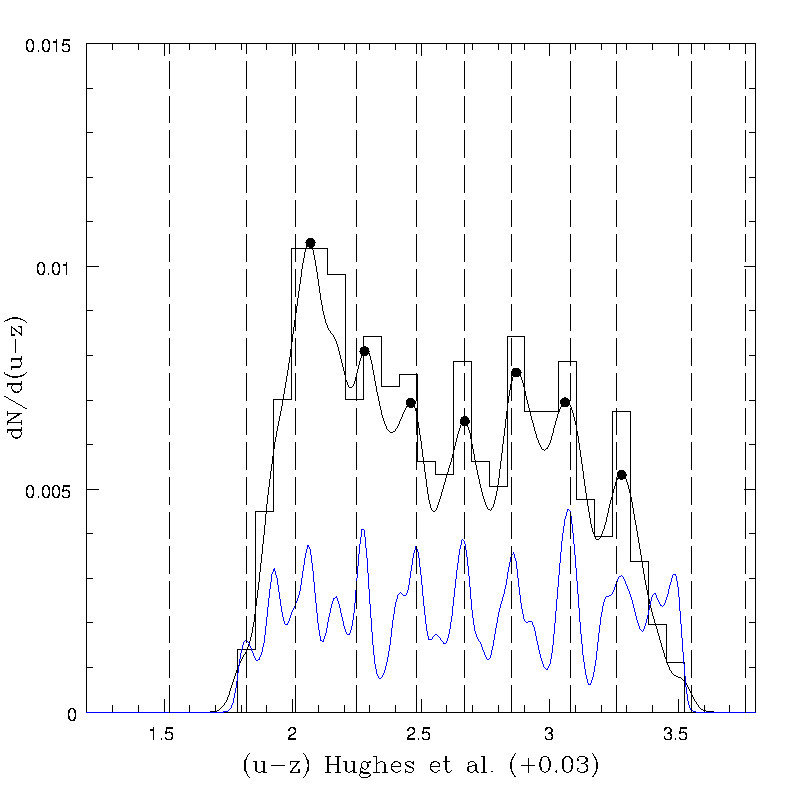}
\vspace{-0.5cm}
    \caption{$(u-z)$ colour distribution for 552 $GC$ candidates with a likelihood $>$ 0.85, $r=$18.0~to~20.5 mag, position angles from 0\degr~to 270\degr, and within 2\degr~from the centre of $NGC~5128$, using data from \citet{Hughes2021}. The smoothed colour distribution was obtained adopting a 0.07 mag Gaussian kernel. Vertical dashed lines correspond to the $TVP$ pattern listed in Table~\ref{table_A3}. The blue line is the output from the $RSS$ routine run on the $GC$ sample included in Fig. 20. The original $NSC$ colours have been transformed to the $SDSS$ system by adding +0.03 mag (see text).  
}
    \label{fig:fig22}
\end{figure}
 These relations, were derived using the 1480 $GC$ candidates inside 2\degr~from the centre of $NGC~5128$, and are
 presented in Fig.~\ref{fig:a8}. The elliptical contours in this figure correspond to a total likelihood larger than 0.50. This last figure also shows, as a reference, a sample of 511 $GCs$ with measured radial velocities, and brighter than $g=$23.0, in $NGC~4486$. 
  
  The $(u-z)$ $GCCD$ for that sample clearly shows seven of the $TVP$ components, with $\sigma_{(u-z)}=$~0.021 mag, as displayed in Fig.~\ref{fig:fig20}. The vertical lines indicate the expected position of the $TVP$ components according to Table~\ref{table_A3}. This diagram also includes the output from the $RSS$ routine after 5$\times$$10^{2}$ random sampling cycles. 
 
 Adopting a limiting magnitude $r=$20.5, decreases the photometric errors, and the sample size to 1080 $GC$ candidates, making the pattern contrast more evident as seen in Fig.~\ref{fig:fig21} (with $\sigma_{(u-z)}=$~0.030 mag).
 
 The most clear defintion of the colour pattern is displayed in Fig.~\ref{fig:fig22}. This case corresponds to 552 objects with a total likelihood parameter $>$ 0.85, $r$ magnitudes from 18.0 to 20.5, and position angles from 0\degr~to~270\degr~(with $\sigma_{(u-z)}=$~0.028 mag).
  
 In all the three last figures, the smoothing kernel size was set to 0.070 mag (taking into account the slope of the $(u-z)$ vs. $(g-z)$ colours relation).
 
 \section{Summary of results}
\label{s7}
 This section gives a description of the main results presented in this paper:
 \begin{enumerate}
 \item  The $RSS$ approach confirms previous results and identifies a multi-peaked structure in the $GCCD$ corresponding to a composite $GCs$ sample in seventy one non-giant galaxies included in the  $ACSVS$.  
\item The combination of the statistical approach given by $OA2016$, and the $ugriz$ fits presented in this paper, produces a reliable sample of $GC$ candidates within $R_{gal}=$60\arcmin~in $NGC~4486$. The output from the $RSS$ routine on this sample, leads to the identification of a clear multi-peaked pattern in an area $\approx$~15 times larger than that covered by the $ACSVS$ data.
\item The $TVP$ pattern is easily detected in the $GCCD$ of cluster candidates with $R_{gal}<$ 5\arcmin. This pattern is also detected in a sub-sample of confirmed $GCs$, with $(g-i)$ colours from \citet{Harris2009}, and radial velocities from \citet{Strader2011}. This result proves that the colour pattern is not a consequence of badly removed field contamination or eventual instrumental effects.
\item It must be stressed that $\bf{none}$ of the thirteen non-giant galaxies whose composite $GC$ sample shows a strong pattern (see Fig.~\ref{fig:fig2} and Fig.~\ref{fig:fig3}), is located within $R_{gal}=$5\arcmin. Most of these moderately bright galaxies are spread on an area with a radius of 5\degr~from the centre of $NGC~4486$, $i.e.$, the $TVP$ seems present not only in the central regions, but over a large area in the $Virgo~cluster$. Furthermore, the detection of the $TVP$ features in $NGC~5128$, through data and membership analysis completely different to those presented in this paper, indicates that the $TVP$ could be traced to the outskirts of the $Virgo~ Super-cluster$.  
\item According to the numerical models described in Section 2, the probability of a fortuitous coincidence between the $GC$ colour pattern present in the non giant galaxies, and that found for the $NGC~4486$ clusters, being independent data sets, would be in the order of 2.5$\times$$10^{-3}$. This probabilty decreases to 1.25$\times$$10^{-4}$ if the case of $NGC~5128$ is included.
\item A bi-dimensional analysis, in the $(g-i)$ vs $(g-z)$ plane, indicates that the $TVP$ can be detected inside $R_{gal}=$45\arcmin,  i.e., neither sampling noise, nor colour gradients or the presence of an eventual competing pattern, affect its detectability on a large angular scale.
\item The $R_{med}$ and $R_{90}$ parameters show a continuous increase as $GCs$ become bluer. The presence of a clear "glitch", or discontinuity, at $(g-z)$~$\approx$~1.15, argues in favor of a  bimodal scenario as those characteristics are indicative of two different $GC$ (blue ad red) families, with distinctic spatial distributions. This behaviour is coherent with the signature of dissipative collapse in both the halo and bulge sub-popultions. However, the prominent blue peak at $(g-z)=$0.86, similar to the dominant blue peak observed for $GCs$ in Virgo galaxies fainter than $M_{g}=$-19.0 (see Fig.~\ref{fig:fig4}), would be consistent with cluster accretion from these less massive galaxies. Dry mergers, in particular, would preserve the $TVP$ features, as they seem common to most galaxies. 
\item Even though, individually, each of the $TVP$ components seem to have null or very mild colour gradients, these gradients can arise when, in a given galactocentric range, the $GC$ sample includes cluster sub-populations with different spatial scale lengths. For example \citet{Forbes2018}, have already noticed that, in some cases, blue $GCs$, if considered as a $\bf{single}$ sub-population, exhibit detectable colour gradients. These colour trends can be explained, as shown in Fig.~\ref{fig:fig18}, by the composition of the diferent blue sub-populations, that reach increasingly larger galactocentric radii as their colors become bluer.
\item Spectroscopic analysis of the $GC$ chemical gradients in $NGC~4486$ has been presented by \citet{Villaume2020}, who find a marked spread of the $[Fe/H]$ index over a large range of galactocentric radius, and also,''remarkably flat'' galactocentric gradients for both blue and red cluster populations. In turn, \citet{Ko2022}, obtain a mild gradient for red $GCs$, and a slightly steeper one characterizing the blue $GCs$ population. These works have distinct $[Fe/H]$ scales but, in both cases, the mean difference between red and blue clusters is $\approx$0.75 dex.

 The clarification of the complex colour structure seen in the $NGC~44486$ clusters, in spectroscopic terms, will require more accurate metallicty determinations, as well as larger sampling volumes. For example, the total population brighter than $i=$22.0 (or  $g$~$\approx$ 23.0), inside $R_{gal}=$45\arcmin,~ is $\approx$ 2800 $GCs$. Random sub-samplings of this population, including ten to twenty percent of these clusters (comparable to the size of the spectroscopic samples), leads to a rather unstable behaviour of the derived galactocentric colour gradients, and presumably, of the chemical abundances.
 
\item The total $GC$ population with $R_{gal}<$ 60\arcmin, assuming a fully Gaussian integrated luminosity function amounts to 14900~$\pm$~1100. This last uncertainty is related with those of the $\alpha=$0.35$\pm$0.05 and $a=$0.030$\pm$0.01 parameters, as well as that of the completeness within the photometric acceptability boundary, that ranges from 0.80 to 0.85.

 This estimate of the total $GCs$ population, derived on the basis of the identification of individual $GC$ candidates, is in very good agreement with \citet{Durrell2014}, who find 14520 $\pm$~1200 $GCs$, through a different approach, based on $GC$ areal density profiles of both blue and red $GCs$. 
\item The complex stream-like structures seen in Fig.~\ref{fig:fig10} deserve further atention in order to confirm that they are the result of merger events as those described by \citet{Romanowsky2012}. We note that $GCs$ associated with feature $\bf{a}$, a confirmed stream, displays all the $TVP$ colour peaks, as well as the other three similar structures ($\bf{b,c,d}$). According to those last authors, streams of this type would be detectable over a time span of about 1 $Gy$, i.e., they formed much later than the events that presumably imprinted the $TVP$ on the whole $GC$ population.
\item The determination of the intrinsic ''colour size'' of each of the $TVP$ components is still precluded by the photometric errors. A preliminary attempt, presented in $F2017$, explored the possibility of decomposing the $GCCD$ corresponding to those galaxies included in Fig. ~\ref{fig:fig2} and Fig. ~\ref{fig:fig3}, in terms of the ''colour spread function'' (i.e., the colour profile of a mono-chromatic $GC$ population after adding the effects of photometric errors and the smoothing kernel). However, a comparison of the model and observed colour-magnitude diagrams rather suggests a continuous sequence of $GC$ colours, with enhancements (or valleys as a consequence of quenching ?), whose effects are detectable as the peaks of the $TVP$. 
\end{enumerate}
\section{Conclusions}
\label{sec8}
 The existence of a multi-peaked $GCs$ colour pattern in Virgo, strongly suggests that the formation of these clusters is not a phenomenon  restricted to a given galaxy but, at least in part, the consequence of a common and repetitive process working on supra-galactic spatial scales. The coincidence of the colour peaks in different galaxies requires that the chemical enrichment in these systems has proceeded at the same pace (although reaching different maximum values as a function of galaxy mass). 
 
 Definite conclusions about the physical nature of the putative phenomena that would modulate the $GCCD$ (and possibly the chemical abundance distribution of the  stellar populations in general) require a solid clarification of the relation between colours, ages and metallicities. Conflicting results are available in the literature concerning this last issue \citep{Kim2017, Usher2019, Villaume2019, Fahrion2020, Kim2021}.
 
 The tentative approach presented in $F2017$ assumes a bifurcated age-metallicity relation (as that observed in the $MW$) as a
possible way to connect the effect of a given external event on both the blue and red $GCs$ populations. This is an arguable assumption since a clear age-metallicity relation for $GCs$ in other galaxies is not known.

 With that caveat in mind, we note the results presented in this work, indicating that blue peaks seem mirrored by a red peak counterpart with a mean $(g-z)$ colour difference of 0.36 mag. Adopting, for example, the colour-metallicity relation presented by \citet{Villaume2019}, that colour separation would translate to a metallicity difference of $\approx$~0.70 dex. This chemical offset is comparable to that observed between the red and blue $MW$~$GCs$ age-metallicity sequences \citep{Leaman2013, Massari2019}. On this basis, a possible connection between coeval blue and red $TVP$ peaks cannot be dismissed.
 
  The detection of the $TVP$ features in the $GCCD$ of $NGC~5128$ uncovers a connection between the $Centaurus~A~Group$ and the $Virgo~cluster$. This is a challenging situation regarding the identification of the physical phenomenon that would be eventually capable of originating a colour structure that is shared by $GCs$ in galaxies currently separated by $\approx$ 16 $Mpc$.

 The case of $NGC~5128$, also indicates the need to revise the situation of $GCs$ in $Local~Group$ galaxies.
 We note that a multi-peaked chemical abundance structure seems to be present in the inner regions of the $MW$, as reported by \citet{Bensby2017}. That result was based on high resolution spectroscopy of dwarf and sub-giant stars observed during micro-lensing events. Even though the observed sample is small (about a hundred stars), it is remarkable that the multi-peaked chemical distribution remains stable as new data are added from subsequent events observed along the years. 
 
 A number of phenomena, like mergers of galaxies, or of sub-galaxy clusters, inducing star forming bursts, followed by rapid quenching originated by post nuclear $SMBH$ activity at high redshifts, appear as promising mechanisms to elucidate the origin of the $GCs$ colour pattern. The effects of multiple mergers, in a Millennium $TNG$ scenario, have been in fact successful in reproducing some of the characteristics of $GC$ systems \citep{Ramos2020}.
 
 Massive galaxies, as $NGC~4486$, have had presumably complex and distinct histories in terms of interactions and mergers, that would be eventually reflected by their $GCCDs$. However, the $Template~Virgo~Pattern$ emerges as an indicator of a common phenomenon affecting the temporal $GCs$ formation sequence in the $Virgo~cluster$ as a whole. The presence of a comparable pattern in the $Fornax~cluster$, justifies a further study in this last, and in other galaxy clusters, in an attempt to clarify both the nature and the spatial scale of the mechanism leading to the $GCs$ colour modulation.
 
\section*{Acknowledgements}
This work is dedicated to Dr. Alejandro Feinstein, a pioneer in the field of stellar photometry and who, kindly and generously, introduced the author into the world of star cluster research. Thanks are due to Dr. Daniel Carpintero, for providing thoroughly tested random number generators used in the $Monte-Carlo$ models and $RSS$ routines, and to Drs. Favio Faifer and Carlos Escudero for useful comments about the manuscript.\\
  
\bibliographystyle{mnras}
\bibliography{biblio_Forte.bib}
\bsp	
\appendix
\section{ Colour-colour relations, globular cluster candidates identification, and additional figures and tables}
\label{a1}
\subsection{Photometric data sources}
\label{a1.1}
 A brief description of the papers with photometric data relevant to this work, follows:
\begin{enumerate}
\item Integrated colours, based on $ugriz$ magnitudes in the $SDSS$ (DR5) system, have been presented by \citet{Chen2010} for galaxies in the $Virgo cluster$. In this sample, sixty galaxies are brighter than $g=$14.0, and their colours,in a galactocentric radius range from 1.0 to 10.0\arcsec were adopted to define the multicolour reference frame. 
\item \citet{Strader2011} published radial velocities for a sample of 735 $GCs$ associated with $NGC~4486$. These authors also present $gri$ magnitudes, taken from CFHT/Megacam observations by \citet{Harris2009}, but calibrated to the $SDSS$ (DR7) system. 
\item $griz$ Gemini and $(C-T_{1})$ photometry for 521 $GC$ candidates in a $GMOS$ field, $\approx$ 5\arcmin~to the south of the $NGC~4486$ centre, was presented in \citet{Forte2013}. This paper introduced a photometric approach to identify $GCs$ based on pseudo-continuum fits defined by $C-griz$ magnitudes. The same method, adopted in the present work, replaces $C$ magnitudes (Washington system) by $u$ magnitudes as defined in the $SDSS$ system.
\end{enumerate}
\subsection{Colour-colour relations and $ugriz$ pseudo-continuums}
\label{a1.2}
 The zero points of the photometric data given in \citet{Strader2011}, \citet{Forte2013}, and \citet{Oldham2016} were shifted to agree with the \citet{Chen2010} colours, according to the values listed in Table~\ref{table_A1}, and combined to define a multicolour reference frame. This was achieved by means of a bi-dimensional piece-wise analysis of each of the  forty five colour-colour relations that can be defined in terms of the $ugriz$ magnitudes. 

 The linear piece-wise analysis detects two mild changes in $\bf{some}$ colour-colour slopes at $(g-i)=$0.90 and 1.10 [or $(g-z)$= 1.08 and 1.35], and in all cases, the linear fits included three colour segments defined by these values, $i.e.$ $(g-i)$ bluer than 0.90, $(g-i)$ from 0.90 to 1.10, and $(g-i)$ redder than 1.10.
 
 The colour-colour relations were used to derive the $ugriz$ pseudo-continuums, listed in Table~\ref{table_A2} as a function of the $(g-i)$ colour (in 0.001 mag intervals). In this table, $g$ magnitudes were arbitrarily set to $g=$15.0. All the colour-colour relations can be derived from this table.
 \subsection{Globular cluster candidates and field objects}
 \label{a1.3}
 \citet{Oldham2016} present $ugriz$ photometry based on their own treatment of $NGVS$ data for objects in a field centred on $NGC~4486$. This catalogue is used for the analysis presented in this work, after transforming their $ugriz$ colours to the $SDSS$ system. This was achieved in three steps: 
 
 First, by approximating the $NGVS$ magnitudes to those of the $SDSS$ system trough the initial zero point shifts listed in Table 1. These shifts give a first approximation to the colour-colour relations defined in terms of the $ugriz$ magnitudes for both systems. Secondly, adopting the relations between the $NGVS$ and $SDSS$ photometric systems presented by \citet{Ferrarese2012}. 
  
  Finally, small corrections ($u:-0.03$; $g:+0.01$; $r:-0.03$; $i:-0.01$; $z:-0.02$), were applied in order to satisfy all the colour-colour relations mentioned  in the previous subsection. A problem with the zero point of the $u$ magnitudes, for objects with $R_{gal}$~$>$ 10\arcmin, is commented below.
 
 $GC$ candidates were identified by comparing their $ugriz$ magnitudes, with the pseudo-continuum listed in Table~\ref{table_A2}. A similar approach was presented in \citet{Forte2013} although, in the present work, the proximity is defined by the minimum $rms$ of the pseudo-continuum fits. This quantity is a function of apparent magnitude as photometric errors increase for fainter objects.
Magnitude fit residuals for each filter band~$i$~are defined as:
\begin{equation}
\epsilon_{i}= [m_{i}-(m_{i_{table}}+\Delta_{i})].\omega_{i}
\end{equation}
 where each magnitude difference is weighted with $\omega_{i}$, the inverse of the photometric $rms$ values given in $OA2016$, for each magnitude $m_{i}$. In this equation, $\Delta_{i}$ is the mean difference between the observed magnitudes and those in Table~\ref{table_A2}, that define the pseudo-continuum delivering the minimum $rms$ of the $\epsilon_{i}$ residuals.
 
 It must be taken into account that, besides the contribution of photometric errors, the $rms$ values will reflect the effect of possible differences between the reference $ugriz$ pseudo-continuums and those that characterize a given $GC$. These differences may arise as a consequence of different horizontal branch morphologies, ages, or the presence of multiple stellar populations.  

 A quantitative boundary, separating $GCs$ and field objects, was determined by analysing the behaviour of the $rms$ values as a function of the $g$ magnitudes. This was accomplished by means of $GCs$ with radial velocities from $ST2011$, and objects with a probability $P$~$>$ 0.90 of being a genuine blue or red $GC$ with $R_{gal}<$ 10\arcmin.

 The analysis shows that the limiting boundary between $GC$ candidates and field objects, can be approximated as:
\begin{equation}
\delta=a.e^{\alpha.(g-g*)}
\end{equation}
with $\alpha=$0.35$\pm$0.05,~$a=$0.03$\pm$0.005, for $g$~$>$~$g*$, and $\delta=a$ for $g$~$<$~$g*$;  with $g*=$21.5.

 This boundary contains 85 percent (within $R_{gal}=$10\arcmin) to 80 percent (at larger $R_{gal}$) of the $GC$ candidates.
 
\begin{figure}
	\includegraphics[width=\columnwidth]{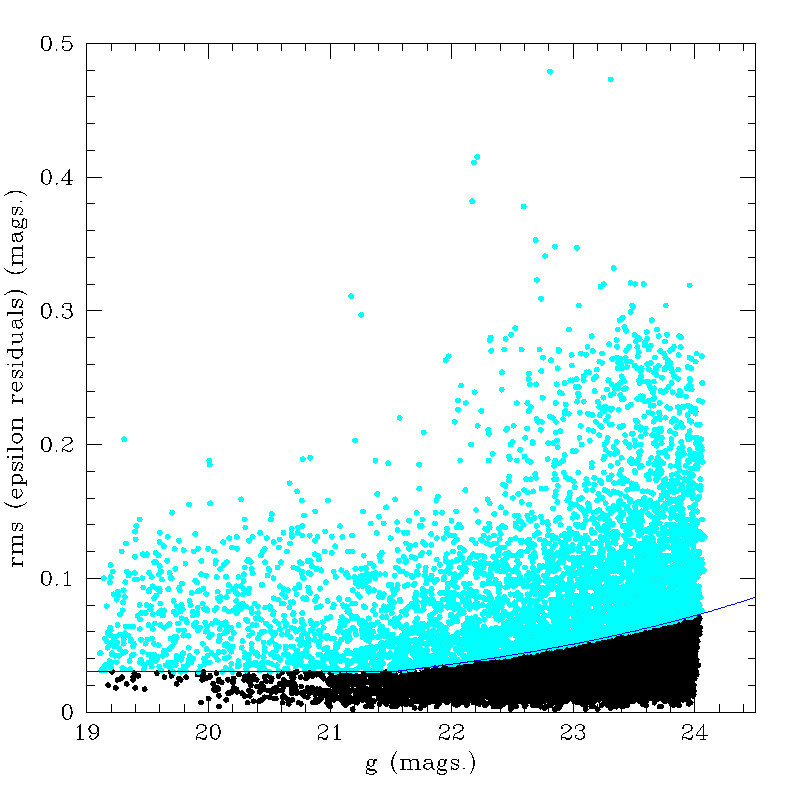}
    \caption{$rms$ of the fit residuals, $\epsilon_{i}$, as a function of $g$ magnitudes for 13271 objects within 60\arcmin~from
    the centre of $NGC~4486$. Black dots are considered as $GC$ candidates. Cyan dots are classified as field objects.
}
    \label{fig:a1}
\end{figure}

  Fig.~\ref{fig:a1} shows the behaviour of the $rms$ of the $\epsilon_{i}$ residuals as a function of $g$ magnitudes, and the domain of both $GC$ candidates (black dots)and field interlopers (cyan dots) for a total of 13172 objects inside $R_{gal}$=60\arcmin.

\begin{figure}
	\includegraphics[width=\columnwidth]{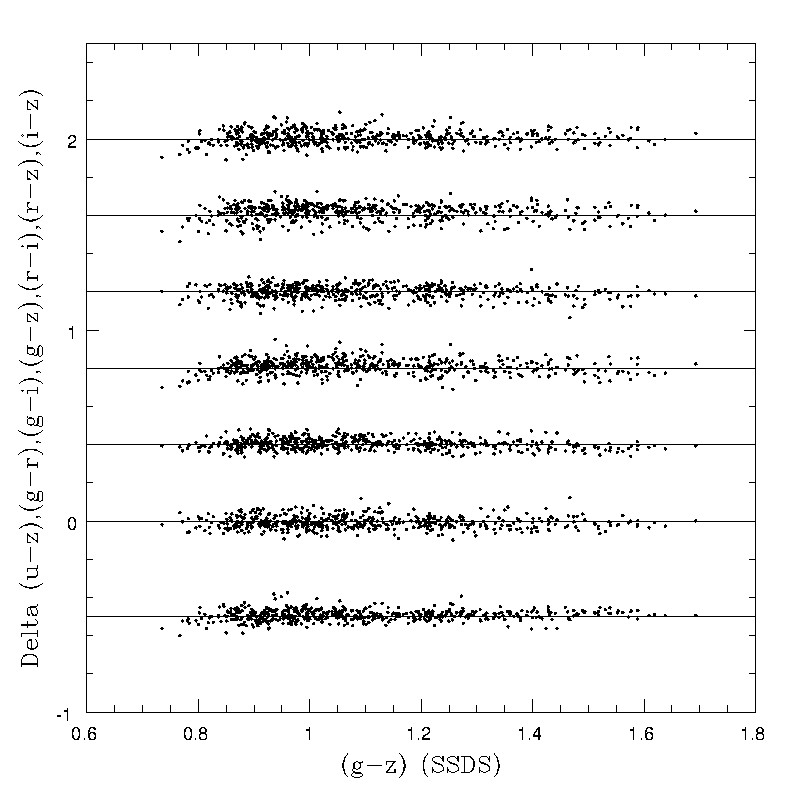} 
    \caption{Colour residuals from the $ugriz$ fits to photometric data of 511 $GCs$ with measured radial velocities as a function of $(g-z)$ colour. Each set of residuals is arbitrarily shifted by 0.2 mag. in ordinates.
}
    \label{fig:a2}
\end{figure}

\begin{figure}
	\includegraphics[width=\columnwidth]{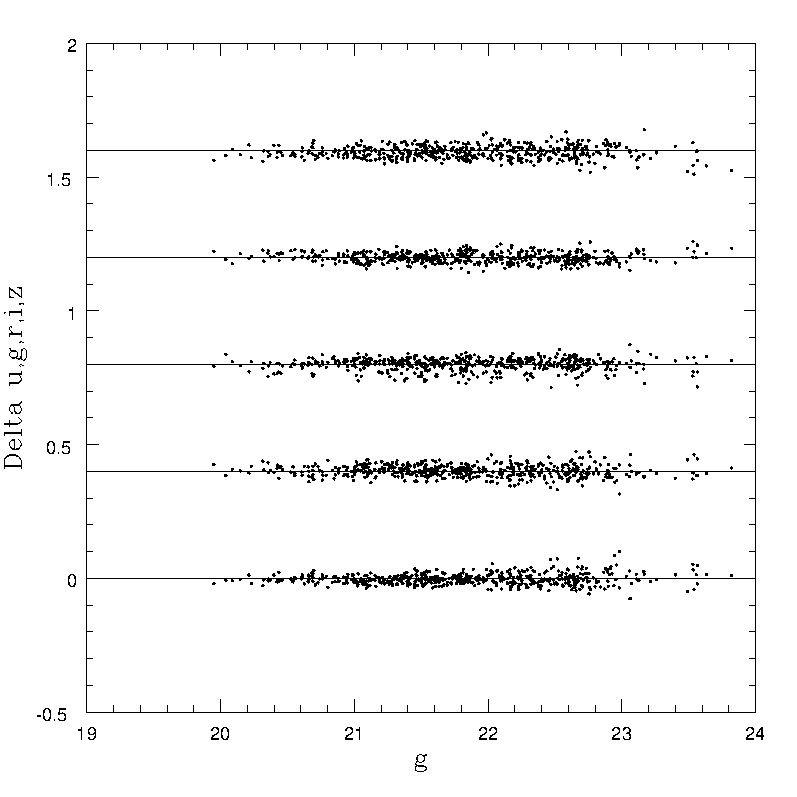}
    \caption{Magnitude residuals from the $ugriz$ fits to photometric data of 511 $GCs$ with measured radial velocities as a function of  $g$ magnitudes. Each set of residuals is arbitrarily shifted in ordinates.
}
    \label{fig:a3}
\end{figure}

 In turn, Fig.~\ref{fig:a2} and Fig.~\ref{fig:a3} display the fit residuals of the $ugriz$ magnitudes, as a function of $(g-z)$ colours and $g$ magnitudes, respectively, for 511 $GCs$ with radial velocities from $ST2011$. These diagrams show no systematic residuals with $GC$ colours, brightness, or galactocentric radii, indicating that the $ugriz$ pseudo-continuums listed in Table~\ref{table_A2}, provide a good representation of the cluster colours. In this process, we noted that $u$ magnitudes, in particular, do show a systematic difference in the sense that $GCs$ with $R_{gal}$~$>$ 10\arcmin~are $0.09$ mag fainter than those in the inner region. 
 
\begin{figure}
	\includegraphics[width=\columnwidth]{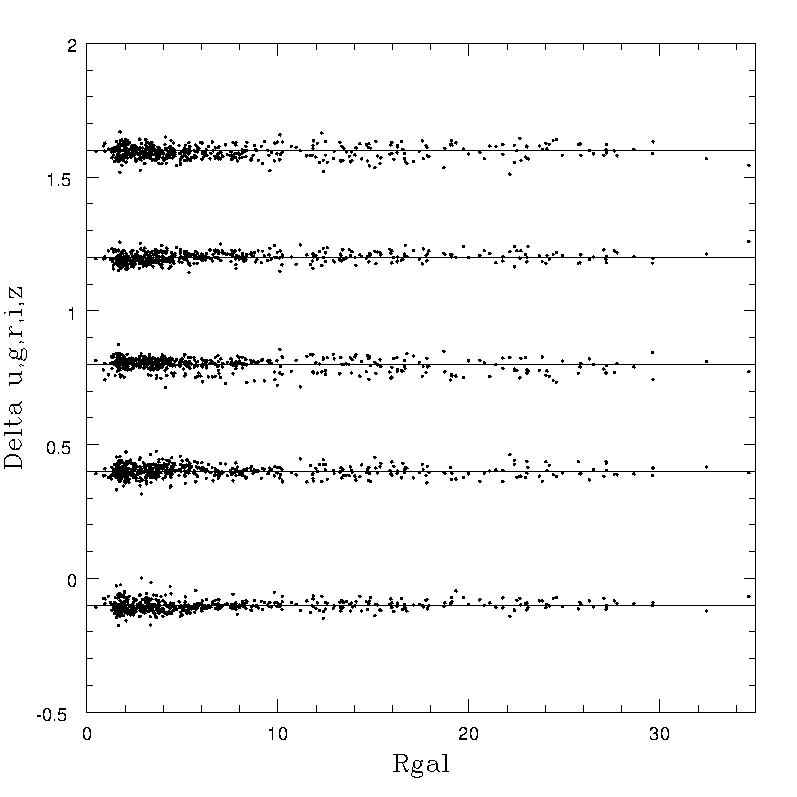}
    \caption{Magnitude residuals from the $ugriz$ fits to photometric data of 511 $GCs$ with measured radial velocities as a function of  galactocentric radii (in arcmin). Each set of residuals is arbitrarily shifted in ordinates.
}
    \label{fig:a4}
\end{figure}

 This difference appears as a marked feature, more compatible with a photometric zero point shift than with the effect of a physical gradient. Ultraviolet magnitudes for objects outside $R_{gal}=$10\arcmin~ were then corrected by $-0.09$ mag. The final magnitude residuals as a function of $R_{gal}$ are shown in Fig.~\ref{fig:a4}.
 
\begin{figure}
	\includegraphics[width=\columnwidth]{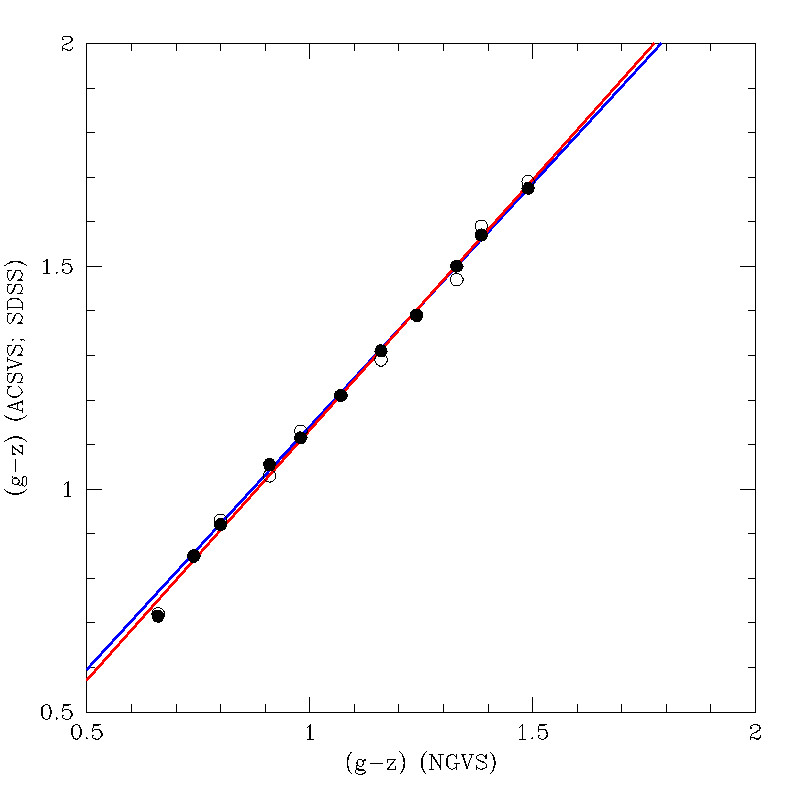}
    \caption{Correlations between the $(g-z)$ colours of the peaks detected in the $SDSS$ system (open dots), and in the $ACSVS$ system (filled dots) with those found on the $NGVS$ data. The blue line is the $SDSS$-$NGVS$ relation presented in Section 3. The red line corresponds to that derived by \citet{Wu2022} (shifted by $+0.015$ mag in ordinates).
}
    \label{fig:a5}
\end{figure}

 Fig.~\ref{fig:a5} shows the relation of the $TVP$ peaks found in the $SDSS$ and $ACSCS$ systems vs. those in the $NGVS$ system. This diagram shows no significant differences between the colours of the peaks detected in the first two systems.
 
\begin{figure}
	\includegraphics[width=\columnwidth]{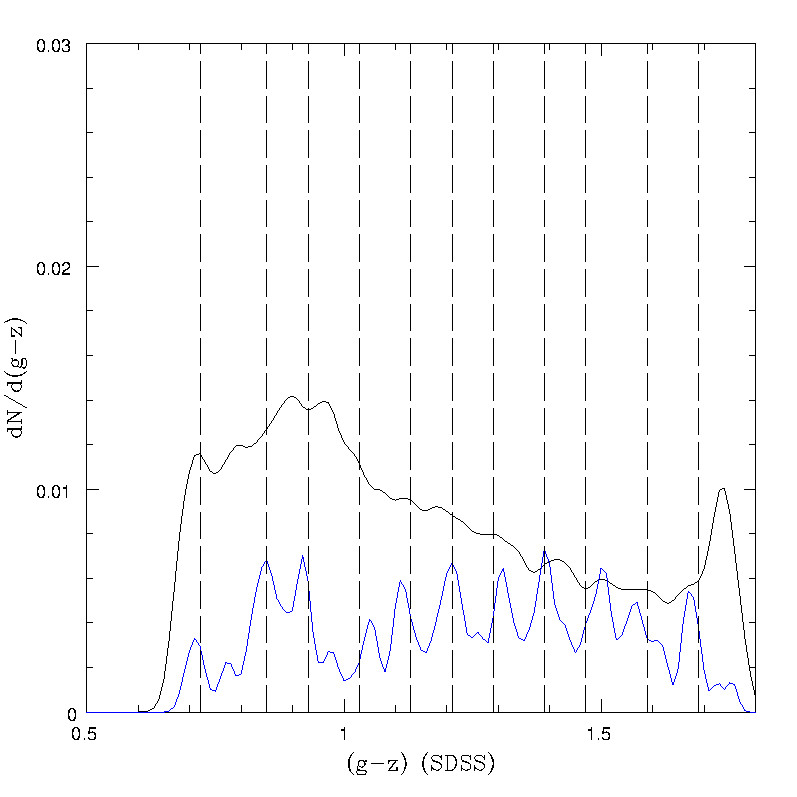}
    \caption{Smoothed $(g-z)$ colour distribution for 6817 objects within 60\arcmin~from the centre of $NGC~4486$ and considered as field interlopers. The blue line is the scaled down output from the $RSS$ routine for $GC$ candidates. Dashed lines correspond to the $TVP$ colours.
}
    \label{fig:a6}
\end{figure}

 Fig.~\ref{fig:a6} depicts the $GCCD$ for 6817 objects considered as field interlopers within 60\arcmin~from the centre of $NGC~4486$.
   
\subsection{GCCD in three glactocentric radius ranges in NGC 4486}
 \label{a1.4}
\begin{figure}
	\includegraphics[width=\columnwidth]{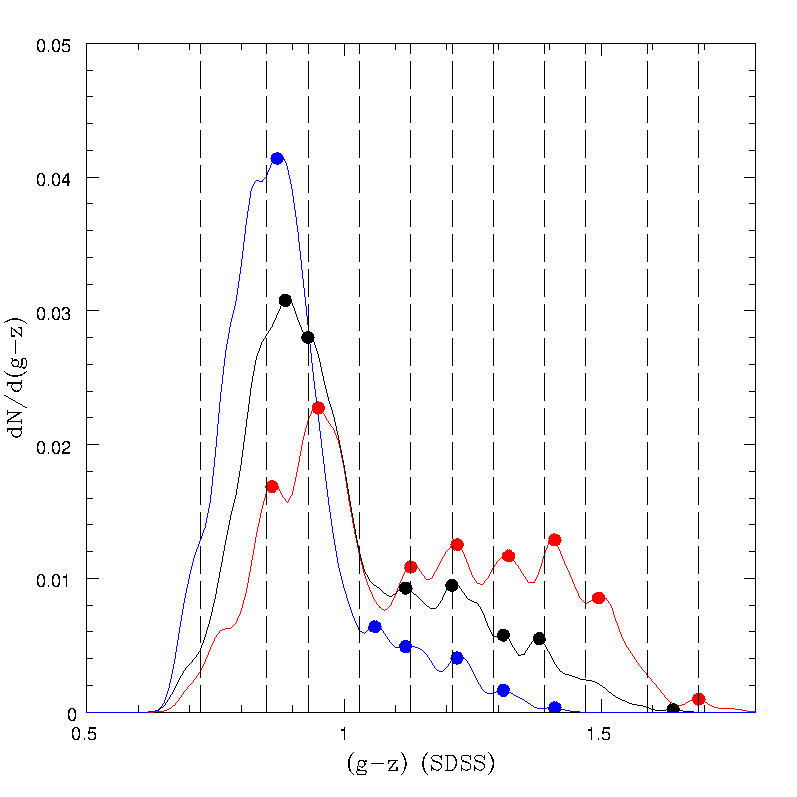}
    \caption{Smoothed $(g-z)$ colour distributions for 3128 $GC$ candidates in three different galactocentric ranges. The red line  belongs to: $R_{gal}=$0.5 to 5\arcmin; black line: 5 to 14\arcmin; and blue line: 14 to 45\arcmin~(see text). Dashed lines correspond to the $TVP$ colours listed in Table~\ref{table_A3}. Red, black and blue dots indicate colour peaks found in each galactocentric sample.
}
    \label{fig:a7}
\end{figure}

 The smoothed $(g-z)$ colour distribution for a total of 3128 $GC$ candidates in three different galactocentric ranges (0.5\arcmin~to 5\arcmin,~5\arcmin~to 14\arcmin~and 14\arcmin~to 45\arcmin) is shown in Fig.~\ref{fig:a7}. 
 
 \subsection{(u-z), (r-z) and (u-r) colour relations for GCs in NGC 5128}
 \label{a1.5}
\begin{figure}
	\includegraphics[width=\columnwidth]{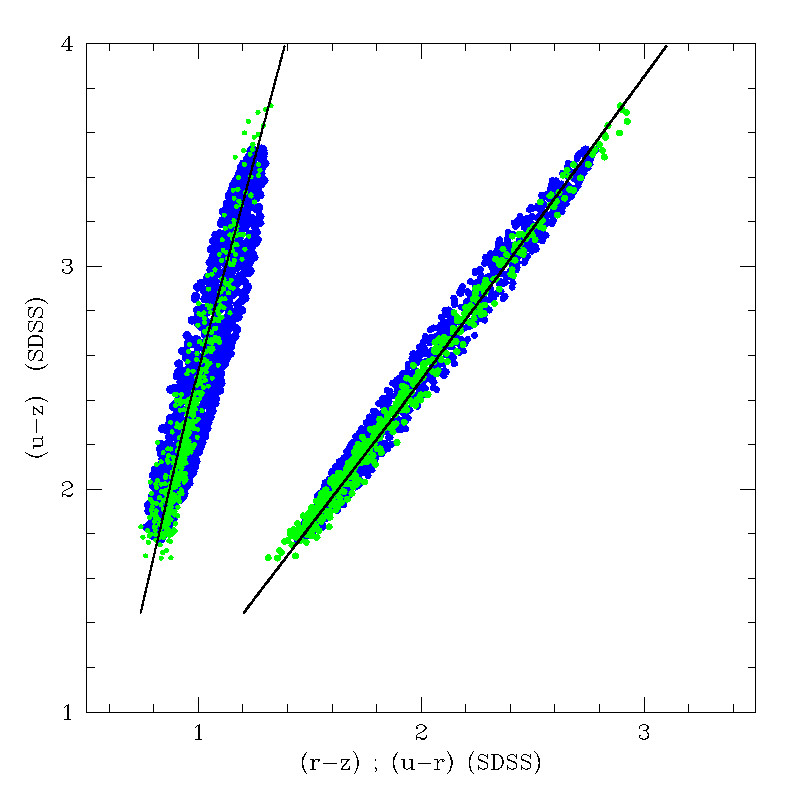}
    \caption{$(u-z)$ colour vs. $(r-z)$ (left) and $(u-r)$ (right) for 1480 $GC$ candidates with a likelihhod larger than 0.50, in $NGC~5128$ (blue dots) from \citet{Hughes2021}. As a reference, green dots are 511 $GCs$ with radial velocities in $NGC~4486$, from Strader et al. (2011). The black lines are the colour relations derived from Table~\ref{table_A2}.  
}
    \label{fig:a8}
\end{figure}

  Fig.~\ref{fig:a8} displays the $(u-z)$ vs. $(r-z)$ and $(u-r)$ colour relations for $GCs$ in $NGC~5128$, using data from \citet{Hughes2021}, once transformed to the $SDSS$ system through zero point shifts (see text).
\newpage
\newpage
\begin{table}
\centering
\caption{Adopted ugriz zero point shifts.}
\begin{tabular}{c c c c c c c c c c c c}
\hline
\hline
\textbf{Source}&&\textbf{u}&&\textbf{g}&&\textbf{r}&&\textbf{i}&&\textbf{z}&\\
\hline
ST2011&&------&& 0.00& &- 0.01&& 0.00&&-----\\
FO2013&&------&& 0.02& &~  0.03&& 0.00&&- 0.06\\
OA2016&&  0.35&& 0.13& &~ 0.02&&  0.03&&~  0.00\\
\hline
\label{table_A1}
\end{tabular}
\end{table}
\begin{table}
\centering
\caption{Globular cluster ugriz pseudo-continuums as a function of the (g-i) colour (complete table 
in the electronic version of the paper).}
\begin{tabular}{c c c c c c c c c c c c}
\hline
\hline
\textbf{(g-i)}& &\textbf{u}& &\textbf{g}& &\textbf{r}& &\textbf{i}& &\textbf{z}&\\
\hline
0.600& &15.759& &15.000& &14.554& &14.400& &14.313\\
0.601& &15.761& &15.000& &14.553& &14.399& &14.311\\
0.602& &15.762& &15.000& &14.553& &14.398& &14.310\\
.....& &......& &......& &......& &......& &......\\
\hline
\label{table_A2}
\end{tabular}
\end{table}
\begin{table}
\centering
\caption{Template Virgo Pattern in three $SDSS$ colour indices.}
\begin{tabular}{c c c c c c c c c c c c}
\hline
\hline
\textbf{(g-i)}& &\textbf{(g-z)}& &\textbf{(u-z)}&\\
\hline
0.62& &0.72& &1.52&\\
0.72& &0.85& &1.82&\\
0.78& &0.93& &2.01&\\
0.86& &1.03& &2.25&\\
0.93& &1.13& &2.48&\\
0.99& &1.21& &2.67&\\
1.05& &1.29& &2.85&\\
1.12& &1.39& &3.08&\\
1.18& &1.47& &3.26&\\
1.27& &1.60& &3.55&\\
1.33& &1.69& &3.76&\\
\hline
\label{table_A3}
\end{tabular}
\end{table}
\label{lastpage}
\end{document}